\documentclass[twocolumn,showpacs,preprintnumbers,amsmath,amssymb,floatfix]{revtex4}
\usepackage{graphicx}% Include figure files
\usepackage{dcolumn}% Align table columns on decimal point
\usepackage{bm}% bold math

\begin{document}

\preprint{APS/123-QED}

\title{Cross sections for proton-induced reactions on Pd isotopes
\\ at energies relevant for the $\gamma$ process}% Force line breaks with \\

\author{I. Dillmann}\email{i.dillmann@gsi.de}
 \altaffiliation[Previous address: ]{Department of Physics, University of Basel, Switzerland; Now at II. Physikalisches Institut, Justus-Liebig-Universit\"at Giessen, and GSI Helmholtzzentrum f\"ur Schwerionenforschung GmbH, Darmstadt, Germany}
\author{L. Coquard}
   \altaffiliation[Present address: ]{Institut f\"ur Kernphysik, Technische Universit\"at Darmstadt, Germany}
\author{C. Domingo-Pardo}
   \altaffiliation[Present address: ]{Instituto de Fisica Corpuscular (IFIC), E-46071 Valencia, Spain}
\author{F. K\"appeler}
\author{J. Marganiec}
  \altaffiliation[Previous address: ]{University of Lodz, Division of Nuclear Physics, Lodz, Poland}
 \altaffiliation[Present address: ]{ExtreMe Matter Institute EMMI, GSI Helmholtzzentrum f\"ur Schwerionenforschung GmbH, Darmstadt, Germany}
\author{E. Uberseder}
 \altaffiliation[Present address: ]{University of Notre Dame, 225 Nieuwland Science Hall, Indiana, USA}
 \affiliation{Karlsruher Institut f\"ur Technologie (KIT), Campus Nord, Institut f\"ur Kernphysik, Postfach 3640, D-76021 Karlsruhe, Germany}

\author{U. Giesen and A. Heiske}
 \affiliation{Physikalisch-Technische Bundesanstalt (PTB), Bundesallee 100, D-38116 Braunschweig, Germany }

\author{G. Feinberg}
 \affiliation{Soreq Nuclear Research Center, Yavne, Israel}

\author{D. Hentschel and S. Hilpp}
\affiliation{Karlsruher Institut f\"ur Technologie (KIT), Campus Nord, Institut f\"ur Nukleare Entsorgung, Postfach 3640, D-76021 Karlsruhe, Germany}

\author{H. Leiste}
\affiliation{Karlsruher Institut f\"ur Technologie (KIT), Campus Nord, Institut f\"ur Angewandte Materialien, Postfach 3640, D-76021 Karlsruhe, Germany}

\author{T. Rauscher}
\author{F.-K. Thielemann}
 \altaffiliation{Alexander von Humboldt fellow at the GSI Helmholtzzentrum f\"ur Schwerionenforschung GmbH, Darmstadt, Germany}
 \affiliation{Department of Physics, University of Basel, Klingelbergstrasse 82, CH-4056 Basel, Switzerland}

\date{\today}% It is always \today, today,

\begin{abstract}
Proton-activation reactions on natural and enriched palladium samples were investigated via the activation technique in the energy range of $E_p$=2.75~MeV to 9~MeV, close to the upper end of the respective Gamow window of the $\gamma$ process. We have determined cross sections for $^{102}$Pd$(p,\gamma)$$^{103}$Ag, $^{104}$Pd$(p,\gamma)$$^{105}$Ag, and $^{105}$Pd$(p,n)$$^{105}$Ag, as well as partial cross sections of $^{104}$Pd$(p,n)$$^{104}$Ag$^g$, $^{105}$Pd$(p,\gamma)$$^{106}$Ag$^m$, $^{106}$Pd$(p,n)$$^{106}$Ag$^m$, and $^{110}$Pd$(p,n)$$^{110}$Ag$^m$ with uncertainties between 3\% and 15\% for constraining theoretical Hauser-Feshbach rates and for direct use in $\gamma$-process
calculations.
\end{abstract}

%\keywords{Suggested keywords}
\maketitle

\section{\label{intro}Introduction}
Astrophysical models can explain the origin of most nuclei beyond the iron group as a combination of processes involving neutron captures on long ($s$ process) or short
($r$ process) time scales \cite{bbfh57, lawi01}. However, 32 proton-rich stable isotopes between $^{74}$Se and $^{196}$Hg cannot be formed by these neutron capture reactions because they are either shielded by stable isotopes from the $r$-process decay chains or lie outside the $s$-process reaction sequence. These isotopes are attributed to a so-called "$p$ process", and are 10 to 100 times less abundant than the neighboring $s$ and $r$ nuclei.

The astrophysical details of the $p$ process are still under discussion and it is conceivable that several different nucleosynthesis processes and sites may conspire to produce all of the $p$ nuclei. The currently favored astrophysical site is explosive burning in core collapse supernovae, where a shock front heats the O/Ne shell of the
progenitor star to temperatures of 2--3~GK, causing photodisintegration of pre-existing seed nuclei \cite{woho78,woho90,raar95,rahe02}. The seed nuclei are partly already
present in the proto-stellar cloud from which the star formed and are partly created in the weak $s$ process during hydrostatic burning preceding the explosion. The very rare $p$ nuclei $^{138}$La and $^{180}$Ta$^{m}$ cannot be made in this manner but rather originate from neutrino-induced reactions ($\nu$ process \cite{WHH90,HKH05}).

Such a synthesis of proton-rich nuclei by sequences of photodissociations and $\beta^+$ decays is also termed "$\gamma$ process" \cite{woho78,rahe02}. The reaction sequences start with ($\gamma,n$) reactions at stability, producing proton-richer, unstable nuclei which, in turn, are further disintegrated. When ($\gamma,p$) and ($\gamma,\alpha$) reactions become comparable to or faster than neutron emission within an isotopic chain, the reaction path is deflected and feeds chains with lower charge number $Z$. The decrease in temperature at later stages of the $p$ process leads to freeze-out via neutron captures and $\beta^+$ decays, resulting in the typical $p$-process abundance pattern with maxima at $^{92}$Mo ($N$=50) and $^{144}$Sm ($N$=82).

Calculations based on the $\gamma$ process concept can produce the bulk of the $p$ nuclei \cite{raar95,ray90,rahe02}. However, the most abundant $p$ isotopes, $^{92,94}$Mo and $^{96,98}$Ru, as well as the whole region $A$$<$124 are notoriously underproduced. Additionally, the mass region 150$\leq$$A$$\leq$165 also seems underproduced in modern, self-consistent models \cite{rau95,rahe02}. It is not yet clear whether the observed underproductions are due to deficiencies in the astrophysical modeling or in the nuclear physics input. The lack of seed nuclei for the production of the light $p$ nuclei indicates the former for the lighter mass region but reliable nuclear physics input is important in both regions. Alternatives to the core-collapse supernova scenario (e.g., explosion of a mass-accreting white dwarf \cite{home91} or sub-Chandrasekhar mass white dwarf \cite{GJH02}, rapid proton captures in X-ray bursts \cite{scha98,scha01} and in the $\nu p$ process \cite{FML06}) still suffer from insufficient modeling, constraints from meteoritic data \cite{Dau03}, and their uncertain contribution to the total galactic nucleosynthesis. In any case, reliable reaction rates are instrumental for developing a consistent $p$-process picture.

From the mere size of the nuclear reaction network for the $p$ process, which includes about 1800 isotopes and more than ten thousand reactions mostly with unstable nuclei, it becomes obvious that the vast majority of reaction rates has to be determined theoretically. 
%Nevertheless, accurate experimental data are important in several respects, either for testing theoretical predictions or for direct use in actual $p$ process network calculations.

\begin{figure*}[!htb]
\includegraphics[scale=0.9]{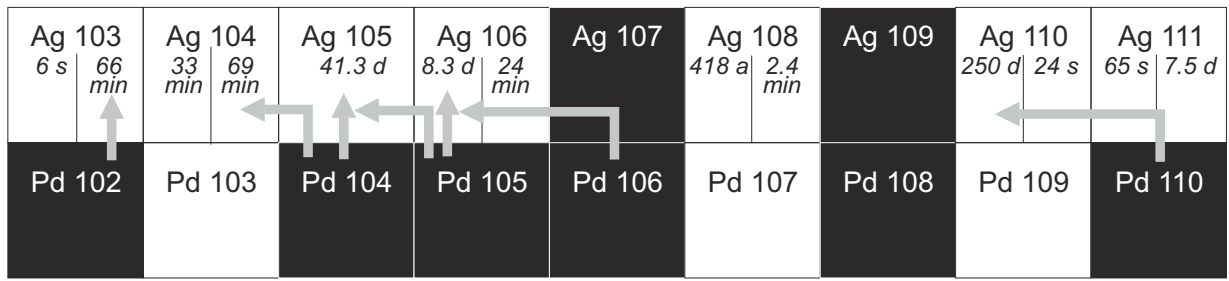}
\caption{\label{Pd-chain}Isotopic chains of Pd and Ag with the investigated reactions.}
\end{figure*}

The experimental database for the $p$ process is -- despite many efforts in the last years -- still very limited, because measurements of the small cross sections of charged-particle reactions below the Coulomb barrier represent a continuing experimental challenge. In this work we present the results of ($p,\gamma$) and ($p,n$) reaction studies on several Pd isotopes (as indicated in Fig.~\ref{Pd-chain}) close to the astrophysically relevant energies of the $p$ process \cite{TR10}.

The importance of the measured ($p,\gamma$) reactions for the $p$ process is twofold. Firstly, theoretical predictions for their cross sections can be tested in order to improve the nuclear reaction modeling for the $\gamma$ process. Secondly, reaction rates derived from the measurements can directly be included in reaction networks for $p$ nucleosynthesis. 

The deflections in the $\gamma$-process path are governed by ($\gamma,p$) reactions in the lighter mass range \cite{Rau06}. Although the ($\gamma,p$) rate will dominate over ($p,\gamma$) in the $\gamma$ process, it has been shown that it is always more advantageous to measure the capture rate and derive its inverse rate by application of detailed balance \cite{KRG08,RKG09}. This is because the reaction rate has to include the thermal population of excited target states in the astrophysical plasma which leads to cross section modifications relative to the cross section of the reaction proceeding only via the ground state of a target nucleus. Since only the latter can be studied in the laboratory it is desirable to measure in the direction of least alteration due to stellar plasma effects. The stellar enhancement factor $\mathrm{SEF}=r^*/r^\mathrm{lab}$ is defined as the ratio of the stellar rate $r^*$, including reactions from thermally populated states of the target nucleus, and the laboratory rate $r^\mathrm{lab}$ with reactions proceeding only from the ground state of the target nucleus \cite{Ili07}. For the $p$ process, the SEF is always smaller for capture than for photodisintegration \cite{UMZ06,RKG09}.

The importance of ($n,p$) reactions for the lower mass range of $p$ nuclei has been pointed out in \cite{RGW06}. For proton-rich nuclei, ($n,p$) reactions have a positive reaction $Q$ value and are in general less affected by stellar plasma effects than ($p,n$) reactions. However, it was pointed out recently that the stellar cross section modification can be suppressed by the Coulomb barrier and that the SEF may be lower for some ($p,n$) reactions than for their ($n,p$) counterpart \cite{KRG08,RKG09}. The reaction $^{105}$Pd($p,n$)$^{105}$Ag is such a case among the reactions presented here. Its SEFs are only $1.1-1.0$ in the relevant plasma temperature range of $2.0-3.5$ GK, compared to SEFs of $1.3-2.7$ for its inverse reaction. For the other ($p,n$) reactions presented in this work, we could only determine partial cross sections to the ground or isomeric state. These cannot be directly converted to astrophysical reaction rates but can be used -- with the aid of theoretical calculations -- to test the description of the proton optical potential, which is also essential in the prediction of the capture and photodisintegration rates.

We commence by describing the experimental technique and sample preparation in Sec.~\ref{exp}, followed by the data analysis (Sec.~\ref{data}), and the experimental results in Sec.~\ref{res}.

\section{Experimental technique}\label{exp}
All cross section measurements have been carried out at the cyclotron and Van de Graaff accelerator of Physikalisch-Technische Bundesanstalt (PTB) in Braunschweig/ Germany \cite{BCD80} with the activation technique by irradiation of thin sample layers and subsequent counting of the induced activity. The Van de Graaff accelerator was used for energies up to 3.5~MeV, above that energy up to 9.0~MeV the cyclotron was used. The astrophysically relevant energy ranges for the measured reactions at temperatures of 2 and 3~GK are listed in Table~\ref{tab:gamow}. In most of the measurements, our data reaches into the energy window relevant in the $\gamma$ process.

\begin{table}[!htb]
\caption{\label{tab:gamow} Astrophysically relevant energy windows for the measured $^A$Pd$(p,\gamma)$ and $^A$Pd$(p,n)$ reactions at 2 and 3~GK \cite{TR10}. Values in brackets refer to the range at 3~GK. The last column shows the measured energy range.}
\renewcommand{\arraystretch}{1.1} % enlarge line spacing
\begin{ruledtabular}
\begin{tabular}{ccccc}
Reaction  & $T$ 	 & Lower end & Upper end & Measured \\
					& (GK) & (MeV)		 & (MeV)		& (MeV) \\
\hline
$^{102}$Pd$(p,\gamma)$ & 2 (3) & 1.64 (2.04) & 3.30 (4.29) & 2.68$-$6.85\\													
$^{104}$Pd$(p,\gamma)$ & 2 (3) & 1.67 (2.13) & 3.40 (4.45) & 2.69$-$5.04\\
$^{105}$Pd$(p,\gamma)$ & 2 (3) & 1.51 (1.74) & 2.81 (3.90) & 2.71$-$5.04\\
\hline
$^{104}$Pd$(p,n)$ & 2 (3) & 5.07 (5.07) & 5.70 (6.07) & 5.80$-$8.82\\	
$^{105}$Pd$(p,n)$ & 2 (3) & 3.87 (4.24) & 5.50 (5.98) & 2.69$-$5.04\\
$^{106}$Pd$(p,n)$ & 2 (3) & 3.75 (3.76) & 4.35 (4.69) & 3.18$-$4.91\\
$^{110}$Pd$(p,n)$ & 2 (3) & 1.71 (1.80) & 2.92 (3.74) & 3.43$-$8.81\\
\end{tabular}
\end{ruledtabular}
\end{table}

\subsection{Sample preparation}
Samples of natural composition were prepared by sputtering $\approx$400~nm 
thick layers of Pd metal onto 1~mm thick Al disks 35~mm in diameter.  
Enriched $^{104}$Pd and $^{105}$Pd samples (from STB Isotopes, Germany, Table~\ref{tab:abun}) were 
first produced by electrodeposition of a PdCl$_2$ solution on tantalum backings 
(see \cite{TF89} for a sketch of the electrolysis cell), but these layers did 
not properly adhere to the backings. Instead, the samples were made by careful 
and uniform deposition of 100 $\mu$l of the PdCl$_2$ solution within the area 
of the beam spot (12 mm in diameter) and subsequent drying.
 
The thicknesses of the Pd samples were determined by X-Ray fluorescence (XRF) by
irradiation of the samples with the bremsstrahlung spectrum of a rhodium anode. 
The induced characteristic Pd X-rays were analyzed by reflection on a LiF crystal. 
This setup was calibrated relative to a blank sample and to six Pd reference 
samples (50 - 500 $\mu$g) prepared from a standard solution. 

The natural samples were between 420 and 520~nm in thickness, corresponding to a Pd mass of (395$-$490)~$\mu$g or (2.9$-$3.6)$\times$10$^{18}$
atoms/cm$^2$. The enriched samples were about twice as thick and contained 
(680$-$850)~$\mu$g or (3.4$-$4.2)$\times$10$^{18}$~atoms/cm$^2$ of $^{104}$Pd or $^{105}$Pd.

\begin{table}
\caption{\label{tab:abun}Mass fractions (in \%) of the natural \cite{iupac03} and enriched samples.}
\renewcommand{\arraystretch}{1.1} % enlarge line spacing
\begin{ruledtabular}
\begin{tabular}{cccc}
				& Natural sample & \multicolumn{2}{c}{Enriched sample} \\
				& 							& $^{104}$Pd & $^{105}$Pd \\
\hline
$^{102}$Pd & 1.02 (1) & 0.10 (2) & 0.01 (2) \\
$^{104}$Pd & 11.14 (8) & 97.0 (1) & 0.32 (10) \\
$^{105}$Pd & 22.33 (8) & 2.30 (5) & 94.50 (5) \\
$^{106}$Pd & 27.33 (3) & 0.45 (5) & 4.59 (5) \\
$^{108}$Pd & 26.46 (9) & 0.10 (2) & 0.46 (2) \\
$^{110}$Pd & 11.72 (9) & 0.05 & 0.13 \\
\end{tabular}
\end{ruledtabular}
\end{table}

\subsection{Experimental setup}
\begin{figure}
\includegraphics{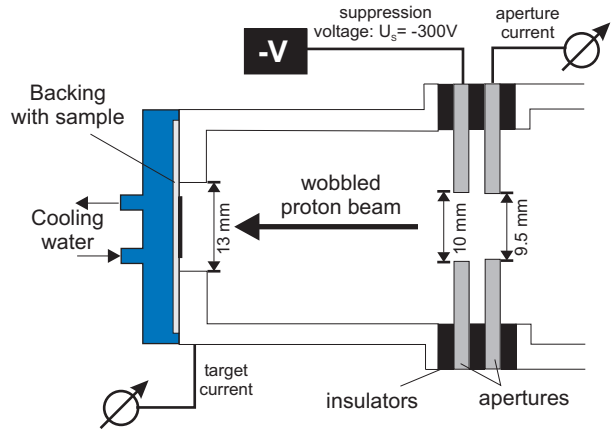}
\caption{\label{setup-ptb}(Color online) Experimental setup at the PTB beamline.}
\end{figure}

The samples were irradiated in an activation chamber, which was designed 
as a Faraday cup (Fig.~\ref{setup-ptb}). The charge deposited on the sample 
was recorded in short time steps by a current integrator for off-line 
correction of beam fluctuations (which turned out to be negligible in the 
end).

The proton beam was wobbled across the chamber aperture indicated in Fig.~\ref{setup-ptb} for homogeneous illumination of the Pd samples. The aperture was slightly smaller than the negatively charged diaphragm (U$_S$=$-$300~V) 
for suppression of secondary electrons. The samples with the Pd layers 
were water-cooled from the outside. For each energy step the proton-beam spot 
was adjusted by means of a quartz window to ensure that the sample was 
completely illuminated. The beam energy was defined within an uncertainty 
of $\pm$25 keV by means of the field calibration of two analyzing magnets
as well as by a time-of-flight measurement of the particle velocity \cite{Boe02}. At energies below 3.5~MeV the Van de Graaff accelerator was used, where the uncertainty of the beam energy calibration was less than 3~keV.

For each activation the effective proton-energy was determined according to 
Eq. 4.99 in \cite{Ili07},  
\begin{equation}
   E^{\rm{eff}}_p=E_{\rm{c.m.}}-\Delta+\Delta \cdot 
\left(-\frac{\sigma_2}{\sigma_1-\sigma_2}+ \left[\frac{\sigma_1^2 + 
\sigma_2^2}{2(\sigma_1-\sigma_2)^2}\right]^{1/2}  \right) \label{eq:eff}
\end{equation}
where $\Delta$ is the target thickness (energy loss) calculated with the Monte Carlo program SRIM 2003 \cite{SRIM03}, $E_{\rm{c.m.}}$ the respective center-of-mass energy, and $\sigma_1$, $\sigma_2$ the measured cross sections of two neighboring points. As can be seen from Eq.~\ref{eq:eff} 
the error bars of $E^{\rm{eff}}$ become asymmetric, with the smaller component in positive direction due to the correction factor in brackets.
 
The samples were activated at 16 different proton energies between 2.75~MeV and 9.00~MeV switching between short-time activations (up to 7200~s for the 65.7~min ground-state in $^{103}$Ag) and long-time activations (up to 36000~s at 2.75~MeV). The average beam current was 10~$\mu$A for the natural samples and 5 $\mu$A for the enriched samples.

\begin{figure*}[!htb]
\includegraphics{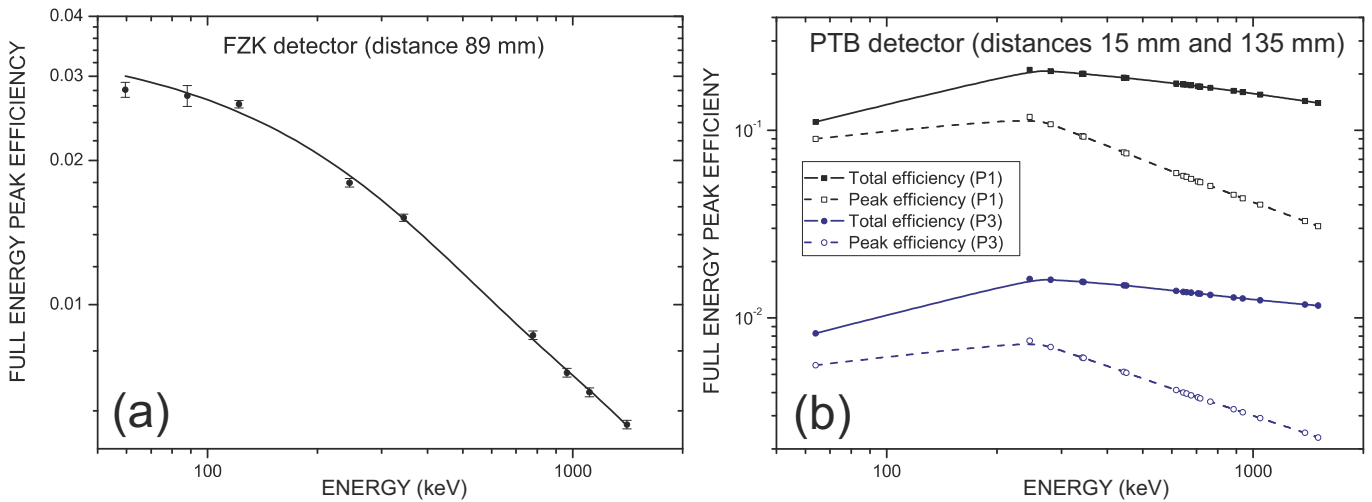}
\caption{\label{eff-ptb}Efficiency curves of the two HPGe detectors: (a) FZK detector, (b) PTB detector. The measured efficiency values of the PTB detector are connected with lines to guide the eye.}
\end{figure*}

The produced activity was measured off-line with two different HPGe detectors (efficiency curves in Fig.~\ref{eff-ptb}), which were shielded from room background by 10~cm lead. The efficiencies shown in Fig.~\ref{eff-ptb} were determined with an uncertainty of 2\% by a set of calibrated reference sources. Apart from the 64 keV transition in $^{105}$Ag, which is very close to the calibration point at 60 keV measured with an $^{241}$Am source, all analyzed $\gamma$-ray lines are in the well-defined part of the efficiency curves above 100 keV (Table \ref{tab:decay-p}).  

\section{Data analysis}\label{data}
The total amount of activated nuclei $N_{a}$ at the end of the irradiation can be deduced from the number of events $C$ in a particular $\gamma$-ray line (Table~\ref{tab:decay-p})
registered in the HPGe detector during the measuring time $t_m$ \cite{beer80}:
\begin{eqnarray}
N_{a} = \frac{C(t_m)} {S~ \varepsilon_\gamma~ 
I_\gamma~(1-e^{-\lambda~t_m})~e^{-\lambda~t_w}} \label{eq:Z}.
\end{eqnarray}
The factor $t_w$ corresponds to the waiting time between irradiation and activity measurement. The factors $\varepsilon_\gamma$ and $I_\gamma$ account for the HPGe efficiency and the relative $\gamma$ intensity per decay (Table~\ref{tab:decay-p}) of the respective transition, and $S$ is the correction factor for coincidence summing. 

\begin{table}
\caption{\label{tab:decay-p}Decay properties of the product nuclei \cite{nds103,nds104,nds105,nds106a,nds110}. Shown here are only transitions used for analysis. }
\renewcommand{\arraystretch}{1.1} % enlarge line spacing
\begin{ruledtabular}
\begin{tabular}{cccc}
Isotope & t$_{1/2}$ & $E$$_\gamma$ & $I$$_\gamma$ \\
        &  & (keV) & (\%) \\
\hline
$^{103}$Ag$\rm ^{g+m}$  & 65.7 (7)~m    & 118.7 & 31.2 (7)  \\
            &               & 148.2 & 28.3 (5) \\
            &               & 243.9 & 8.5 (5)  \\
            &               & 266.9 & 13.4 (4) \\
            &               & 1273.8 & 9.4 (3) \\
\hline
$^{104}$Ag$^g$  & 69.2 (7)~m    &  740.5 &  7.2 (9)\footnotemark[1] \\
           &               &  758.7 &  6.4 (8)\footnotemark[1]  \\
           &               &  767.6 &  65.7 (21) \\
           &               &  942.6 &  25.0 (23)\footnotemark[1] \\
           &                     & 1625.8 & 5.1 (7)\footnotemark[1] \\
\hline
$^{105}$Ag$^g$  & 41.29 (7)~d   & 64.0  & 10.5 (10) \\
            &               & 280.4 & 30.2 (17)  \\
            &               & 344.5 & 41.4 (6) \\
            &               & 443.4 & 10.5 (5) \\
\hline
$^{106}$Ag$^m$  & 8.28 (2)~d  & 451.0  & 28.2 (7) \\
           &               & 717.2  & 28.9 (8)  \\
           &               & 748.4  & 20.6 (6) \\
           &               & 1045.8 & 29.6 (10) \\
\hline
$^{110}$Ag$^m$  & 249.76 (4)~d & 657.8  & 94.3 (3) \\
            &               & 763.9  & 22.62 (21) \\
            &               & 884.7  & 72.7 (4) \\
            &               & 937.5  & 34.2 (6) \\
\end{tabular}
\end{ruledtabular}
\footnotetext[1]{New intensities, see Table~\ref{104Ag}.}
\end{table}

Throughout the activations the proton flux was recorded in time steps of 1 min,
but time variations were found to be negligible. Therefore, the correction for the decay during activation, $f_b$, could be calculated by the expression for constant flux, $f_b=\frac{1-exp(-\lambda~t_a)}{\lambda~t_a}$. Any $\gamma$-ray self-absorption is negligible due to the thin Pd layers. The cross section (in barn) at the respective proton energy can then be calculated via
\begin{eqnarray}
\sigma(E_p) = \frac{N_{a}~\lambda~t_{a}~10^{24}}{H~N~\Phi_{\rm{tot}}~(1-e^{-\lambda~t_{a}})}. \label{sigma-p}
\end{eqnarray}
$H$ is the abundance of the respective Pd isotope (Table~\ref{tab:abun}), $\Phi_{\rm{tot}}$ the collected proton charge during the activation time $t_{a}$, and $\lambda$ the decay constant. $N$ is the area density of the Pd samples in cm$^{-2}$. 

\subsection{Coincidence-summing corrections}\label{coin}
Coincidence summing occurs when two or more $\gamma$ rays are recorded within the resolving time of the detector \cite{DeHe}. The induced activities were measured with two different HPGe detectors (labelled "FZK detector" and "PTB detector", Fig.~\ref{eff-ptb}). The FZK detector is a $n$-type coaxial detector with a thin carbon window and a crystal volume of 370~cm$^3$ corresponding to a relative efficiency of 100\%. The PTB detector is a $p$-type coaxial detector with an aluminium window and a crystal volume of 300 cm$^3$ corresponding to a relative efficiency of 70~\%. All efficiencies are given with respect to 3"$\times$~3" NaI(Tl) detectors. 
 
Because the FZK detector was only used for measurements at a distance of 89~mm, the summing corrections are low and were estimated with one sample measured also at a larger distance of 164 mm. The respective results were in perfect agreement with the calculated summing corrections. For the PTB detector, which was used at two distances (P1= 15~mm and P3= 135~mm) total and peak efficiencies were available from previous experiments (Fig.~\ref{eff-ptb}) and could be used for the calculation of coincidence summing corrections. In this work, summing corrections are significant only for measurements with the PTB detector at the short distance P1, which had to be used for the runs at the lowest energies.

\subsection{Uncertainties}\label{error}
The systematic and statistical uncertainties from these measurements are summarized in Table~\ref{unc}.
The energy loss of the proton beam in the Pd layer was calculated with
SRIM 2003 \cite{SRIM03} using the optional tables for range and 
stopping power. The samples thickness of typically 460~nm corresponds
to an average energy loss between 30~keV at 2.5~MeV and 15~keV at 9~MeV 
proton energy. The respective uncertainty of the proton energy is 25 keV 
for the cyclotron and $\le$3~keV for the Van de Graaff. As shown in 
Eq.~\ref{eq:eff} the error bars of $E_p^{\rm{eff}}$ are asymmetric, with 
the smaller component in positive direction. The uncertainty in the 
collection of the proton beam current was determined to be $\leq$1\%.

\begin{table*}[!htb]
\caption{Relative uncertainties for the individual measurements in \%. Values in brackets refer to enriched samples.} \label{unc}
\begin{ruledtabular}
 \begin{tabular}{cccccccc}
Source of uncertainty  & $^{102}$Pd($p, \gamma$) & $^{104}$Pd($p, n$) & $^{104}$Pd($p, \gamma$) & $^{105}$Pd($p, \gamma$) 
                       & $^{105}$Pd($p, n$)      & $^{106}$Pd($p, n$) & $^{110}$Pd($p, n$) \\
                       & $\rightarrow^{103}$Ag   & $\rightarrow^{104}$Ag$^{g}$ & $\rightarrow$$^{105}$Ag  
					   & $\rightarrow^{106}$Ag$^{m}$ & $\rightarrow^{105}$Ag & $\rightarrow^{106}$Ag$^{m}$ 
					   & $\rightarrow$$^{110}$Ag$^{m}$ \\
\hline
Isotopic abundance     & 0.98 & \multicolumn{2}{c}{0.72 (0.10)} & \multicolumn{2}{c}{0.36 (0.05)} & 0.11 & 0.77 \\
Detector efficiency    & \multicolumn{7}{c}{2.0} \\
Beam current integration & \multicolumn{7}{c}{1.0} \\
Sample mass (XRF)      & \multicolumn{7}{c}{1.5} \\
$\gamma$-ray intensity & 1.8--5.9  & 3.2--4.3 & 1.5--9.5   & 2.5--3.4    & 1.5--9.5     & 2.5--3.4 & 0.3--1.8  \\
Statistical error      & 2.5--15   & 0.3--0.6 & 0.2--7.7   & 3.0--19     & 0.2--3.6     & 3.0--12  & 0.7--14  \\
					   & \multicolumn{7}{c}{ } \\
Total uncertainty      & 4.1--16   & 4.8--5.1 & 3.0--13   & 4.7--19     & 3.0--11.0    & 4.7--13  & 2.7--14   \\  \end{tabular}
\end{ruledtabular}
\end{table*}

The emission probabilities of the $\gamma$ transitions in the decay of $^{104}$Ag$^{g}$ exhibit rather large uncertainties (see Table~\ref{tab:decay-p}). Some of these uncertainties could be reduced by factors of 3 by normalization to the strongest transition at 767.6 keV. The previous and improved intensities are compared in Table~\ref{104Ag}

\begin{table}
\caption{Improved $\gamma$-ray intensities (in \%) for the decay of $^{104}$Ag$^{g}$.} 
\label{104Ag}
\renewcommand{\arraystretch}{1.1} % enlarge line spacing
\begin{ruledtabular}
 \begin{tabular}{ccc}
$E$$_\gamma$ (keV) & $I_\gamma$ \cite{nds104} 
		   		 & $I_\gamma$(new)          \\
\hline
740.5            & 7.2 (9)   &    7.19 (31) \\
758.7            & 6.4 (8)   &    6.62 (26) \\
767.6            & 65.7 (21) &   65.7 (21)  \\
942.6            & 25.0 (23) &   23.69 (83) \\
1625.8           & 5.1 (7)   &    5.25 (19) \\
 \end{tabular}
\end{ruledtabular}
\end{table}

At the lower part of the investigated energy range the total uncertainties are
dominated by the poor counting statistics, except for cases 
with a favorable half-life, which could be counted for longer times (e.g.
$^{105}$Ag). The systematic uncertainties are composed of the contributions 
from the detection efficiencies (2\%), the XRF measurement of the sample mass  
(1.5\%), the integrated proton charge (1\%), the $\gamma$-ray intensities 
(Table~\ref{tab:decay-p}), and the isotopic abundances (Table~\ref{tab:abun}) \cite{iupac03}.

\section{Results}\label{res}

In the following, we present our results for each of the reactions individually and compare them to literature data and to published predictions of the NON-SMOKER code \cite{rath00,rath01}. As already explained in Sec.~\ref{intro}, $(p,\gamma)$ and $(p,n)$ reactions and their inverses are directly important in the $\gamma$ process. However, they occur at lower reaction energies than were accessible experimentally. Nevertheless, the data may be compared to theoretical predictions to pinpoint possible systematic problems in the predictions or their input data which also may play a role at lower energy. In this respect, $(p,\gamma)$ and $(p,n)$ reactions contribute different types of information for the treatment of the optical potentials and $\gamma$ strengths. On one hand, the ($p,n$) reactions are most sensitive to the proton strength functions, and thus to the proton optical potential, across almost the total range of measured energies, provided that the averaged neutron widths are dominating the total reaction width. This is the case for energies well above the threshold. A few hundreds of keV above the threshold, neutron widths may be comparable to $\gamma$ and proton widths and the cross section will be sensitive to variations in either of them. On the other hand, depending on the target nucleus, proton capture is more sensitive to the $\gamma$ strengths at higher energies whereas it has a higher sensitivity to the proton optical potential at low energies, where the averaged proton widths become smaller than the averaged $\gamma$ widths. Therefore, an interpretation of possible deviations of theory from experiment has to consider the relative importance of the different nuclear properties at each energy.

The total cross sections for $^{102}$Pd($p, \gamma$)$^{103}$Ag, $^{104}$Pd$(p,\gamma)$$^{105}$Ag, and $^{105}$Pd($p, n$)$^{105}$Ag have been measured in the proton energy range between 2.75 and 9.00~MeV. These results could be complemented by the partial cross sections 
for $^{104}$Pd($p, n$)$^{104}$Ag$^{g}$, $^{105}$Pd($p,\gamma$)$^{106}$Ag$^{m}$, $^{106}$Pd($p, n$)$^{106}$Ag$^{m}$, and $^{110}$Pd($p, n$)$^{110}$Ag$^{m}$.

\subsection{$^{102}$Pd($p, \gamma$)$^{103}$Ag}

The total proton capture cross section of $^{102}$Pd was determined 
via the five $\gamma$ transitions at 118.7, 148.2, 243.9, 266.9, and 
1273.8~keV. The half-life of $^{103}$Ag was also checked and yielded 
68.2 $\pm$ 2.4~min, slightly longer than but still consistent with 
the 65.7 $\pm$ 0.7~min given in \cite{nds103}. The cross sections 
(derived as weighted averages from all five transitions) are 
listed in Table~\ref{tab:102} together with the respective $S$ factors. 

\begin{figure}
\includegraphics{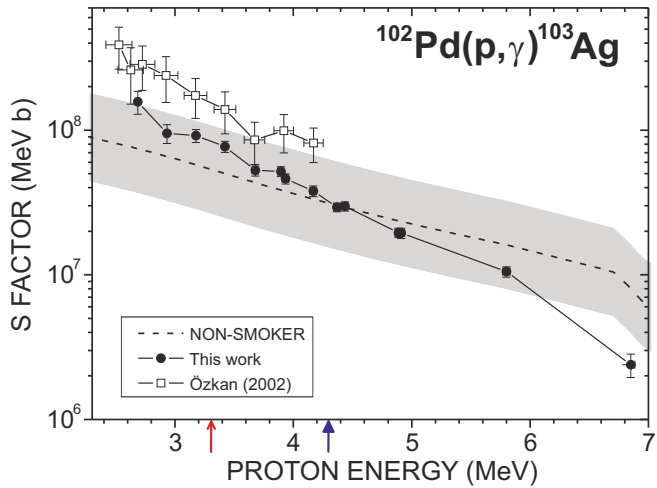}
\caption{\label{pd102-res}(Color online) $S$ factors for $^{102}$Pd($p, \gamma$)$^{103}$Ag. The predictions from NON-SMOKER \cite{rath00,rath01} (dashed line) are plotted with a region of uncertainty of a factor of two. The comparison 
with the $^{102}$Pd($p, \gamma$)$^{103}$Ag results of \"Ozkan {\it et 
al.} \cite{OMB02} (open symbols) exhibits a clear discrepancy. The thin and thick arrows indicate the upper ends of the respective Gamow windows for T=2 and 3~GK, respectively (see Table~\ref{tab:gamow}).}
\end{figure}

The present results are compared in Fig.~\ref{pd102-res} with 
the previous data of \"Ozkan {\it et al.} \cite{OMB02} and with NON-SMOKER 
predictions \cite{rath00,rath01}. The typical factor-of-two uncertainty of 
this prediction for proton-capture reactions is indicated by the 
gray band. Within these limits there is rather good agreement with 
this measurement although the NON-SMOKER values show a somewhat 
different energy dependence. 

The present result are about a factor of three lower than the 
experimental data of \"Ozkan {\it et al.} \cite{OMB02}. These authors 
used enriched $^{102}$Pd samples with an area density of 2~mg/cm$^2$, 
three times thicker than in this work. This explains the 
larger energy spread of the data points (60-90~keV) but does not 
explain the discrepancy between both measurements. 

We have to distinguish three different regions in the energy range studied here. At the lower end up to an energy of about 4.5 MeV the cross section predictions are mostly sensitive to the predicted averaged proton width. Above that energy, the proton width becomes larger than the $\gamma$ width and consequently the sensitivity to the latter dominates. The data point at the highest energy is closely above the $(p,n)$ threshold and the neutron width will have an additional impact there. It can be seen that the energy dependence of the cross section is slightly different in the three different regimes (this is more pronounced in the data) because it is given by the energy dependence of different nuclear properties. Since the proton widths are most important at $p$-process energies, we focus on the lowest energies. The data is higher than the prediction at these energies but it is difficult to identify a trend because of the increasing impact of the $\gamma$ width towards higher energy. The experimental results lie within the expected theoretical uncertainty of a factor of two and are compatible with the assumption that the theoretical proton widths just have to be scaled by a constant factor.

\begin{table}
\caption{\label{tab:102} Cross sections and $S$ factors for 
$^{102}$Pd($p, \gamma$)$^{103}$Ag.}
\renewcommand{\arraystretch}{1.2} 
\begin{ruledtabular}
\begin{tabular}{ccc}
$E{\rm ^{eff}}$    & Cross section     & $S$ factor          \\
   (MeV)                  & (mbarn)           & (10$^{7}$ MeV b) \\
\hline
2.685$^{+0.013}_{-0.018}$ &  0.058$\pm$0.010  &  15.7$\pm$2.8  \\
2.932$^{+0.012}_{-0.018}$ &  0.105$\pm$0.016  &  9.50$\pm$1.4  \\
3.178$^{+0.012}_{-0.016}$ &  0.266$\pm$0.028  &  9.16$\pm$0.95 \\
3.422$^{+0.012}_{-0.015}$ &  0.522$\pm$0.046  &  7.69$\pm$0.68 \\
3.679$^{+0.025}_{-0.029}$ &  0.797$\pm$0.070  &  5.29$\pm$0.47 \\
3.894$^{+0.026}_{-0.029}$ &  1.43$\pm$0.11    &  5.19$\pm$0.41 \\
3.934$^{+0.026}_{-0.029}$ &  1.42$\pm$0.12    &  4.62$\pm$0.37 \\
4.169$^{+0.027}_{-0.028}$ &  2.12$\pm$0.18    &  3.80$\pm$0.32 \\
4.369$^{+0.028}_{-0.029}$ &  2.60$\pm$0.18    &  2.93$\pm$0.20 \\
4.435$^{+0.028}_{-0.030}$ &  3.06$\pm$0.23    &  2.98$\pm$0.22 \\
4.889$^{+0.030}_{-0.030}$ &  5.08$\pm$0.38    &  1.96$\pm$0.15 \\
4.912$^{+0.028}_{-0.032}$ &  5.26$\pm$0.43    &  1.94$\pm$0.16 \\
5.800$^{+0.031}_{-0.033}$ &  12.3$\pm$1.0     &  1.05$\pm$0.09 \\
6.851$^{+0.035}_{-0.037}$ &  10.6$\pm$2.0     &  0.24$\pm$0.04 \\ 
\end{tabular}
\end{ruledtabular}
\end{table}

\subsection{$^{104}$Pd($p, n$)$^{104}$Ag$^{g}$} \label{d104pn}

The $^{104}$Pd($p, n$)$^{104}$Ag reaction channel opens at $E_p$= 5.11~MeV.
The 5$^+$ ground state ($t_{1/2}$= 69.2~min) and the 2$^+$ isomeric 
state ($t_{1/2}$= 33.5~min) of $^{104}$Ag are decaying both via 
$\beta^+$ decay and electron capture (EC) and are not directly 
connected by internal decay. The shorter half-life of the isomer 
made it difficult to derive the isomeric cross section. Therefore,
only the cross section to the ground-state was determined after 
an appropriate waiting time. To avoid interferences, the analysis was limited to those $\gamma$ transitions, where feeding from the isomer is excluded or negligibly weak, i.e. the $\gamma$ lines at 740.5, 
758.7, 767.6, 942.6, and 1625.8~keV. The weighted cross sections 
and the $S$ factors are summarized in Table~\ref{tab:104pn}. 

\begin{figure}
\includegraphics{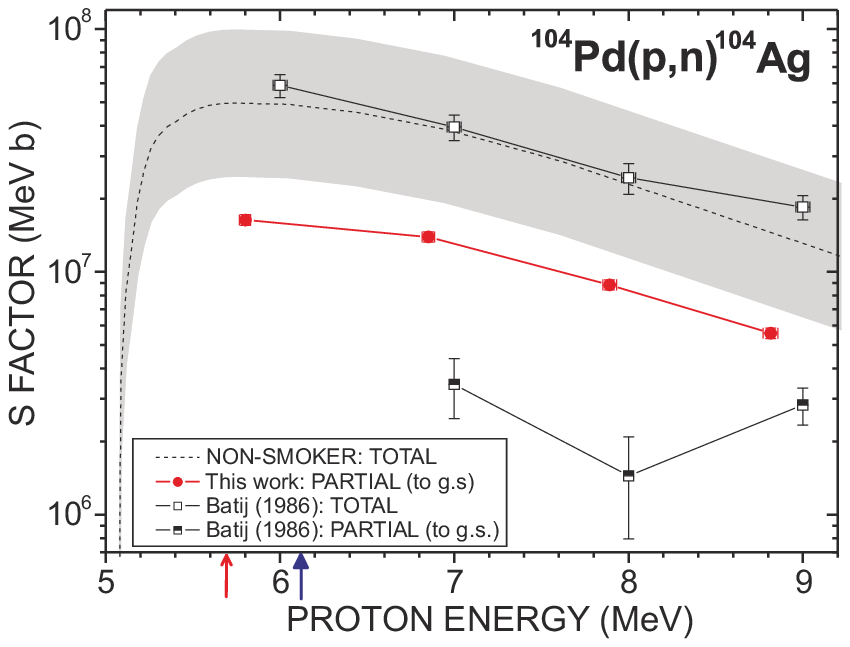}
\caption{\label{pd104pn-res}(Color online) $S$ factors for $^{104}$Pd($p, n$)$^{104}$Ag. The predictions from NON-SMOKER \cite{rath00,rath01} (dashed line) are plotted with a region of uncertainty of a factor of two. For $^{104}$Pd($p, n$)$^{104}$Ag the present results are compared to the measurement of Batij {\it et al.} \cite{BSR86}. The thin and thick arrows indicate the upper ends of the respective Gamow windows for T=2 and 3~GK, respectively (see Table~\ref{tab:gamow}).}
\end{figure}

The comparison in Fig.~\ref{pd104pn-res} shows significant discrepancies with the partial cross sections reported by Batij {\it et al.} \cite{BSR86}. Since NON-SMOKER results are not available for the partial cross section, the total cross section is plotted instead. In spite of the puzzling partial cross sections, the total cross sections of Ref. \cite{BSR86} are in good agreement with the theoretical result, confirming the averaged proton widths obtained by using the optical potential of \cite{lej80} for this energy range. The present (partial) cross section exhibits the same trend as the NON-SMOKER curve. 
%Accordingly, an isomeric ratio $IR=\frac{\sigma_m}{\sigma_{tot}}= 1- \frac{\sigma_g}{\sigma_{tot}}= 0.61-0.67$ can be estimated for the runs between 5.8 and 8.8~MeV. 

Following the EXFOR database \cite{exfor}, the data from Bitao et al. \cite{BZJ98} is listed as the ratio of isomeric to ground state cross section, $X=\frac{\sigma_m}{\sigma_{g}}$. These ratios were deduced via the transitions at 556, 768, and 1239~keV occurring in the decay of the isomer as well as of the ground state. Using our experimental data and the total cross section from Ref.~\cite{BSR86}, we also deduced the ratio $X$, yielding values between 3.7 for 5.802~MeV and 2.8 for 8.815~MeV, in clear disagreement with the values of $X$=0.031$-$0.132 for energies between 6.31 and 8.99~MeV given in \cite{BZJ98}. Unfortunately, these authors do not give any information on how their ratios were calculated nor on the cross section data used. 

\begin{table}
\caption{\label{tab:104pn} Partial cross sections and $S$ factors for 
$^{104}$Pd($p, n$)$^{104}$Ag$^{g}$.}
\renewcommand{\arraystretch}{1.2} % enlarge line spacing
\begin{ruledtabular}
\begin{tabular}{ccc}
$E{\rm ^{eff}}$    & Cross section   & $S$ factor \\
   (MeV)      			  & (mbarn) 		& (10$^7$ MeV b) \\
\hline
5.802$^{+0.030}_{-0.035}$ &  19.2$\pm$1.0   &  1.64$\pm$0.08 \\
6.852$^{+0.035}_{-0.037}$ &  62.1$\pm$3.0   &  1.40$\pm$0.07 \\ 
7.892$^{+0.039}_{-0.040}$ &  111.3$\pm$5.4  &  0.88$\pm$0.04 \\
8.815$^{+0.042}_{-0.043}$ &  149.9$\pm$7.4  &  0.56$\pm$0.03 \\
\end{tabular}
\end{ruledtabular}
\end{table}

\subsection{$^{104}$Pd($p, \gamma$)$^{105}$Ag and $^{105}$Pd($p, n$)$^{105}$Ag}
\label{sec:104105}
\subsubsection{Cross sections measured with natural samples}

Because the $^{104}$Pd($p, \gamma$)$^{105}$Ag and $^{105}$Pd($p,n$)$^{105}$Ag reactions are leading to the same product nucleus, only the sum of the two cross sections can be determined above the ($p, n$) threshold at 2.15~MeV. This composite cross section 
\begin{eqnarray}
\sigma^+ ~(N_{104}+N_{105}) = N_{104}~\sigma_{pg} +
N_{105}~\sigma_{pn} \label{eq:sigmaplus}
\end{eqnarray}
was determined using the $\gamma$ transitions at 64.0, 280.4, 344.5, and 443.4~keV. The 7.23~min isomer in $^{105}$Ag could not be resolved due to its short half-life, but this state decays by 99.66\% via internal transitions. Therefore, the 0.34\% electron capture branch has a negligible effect on the total cross section. The results are shown in Table~\ref{sigmaplus104105}.

\begin{table}[!htb]
\caption{Composite cross section and $S$ factor for the $^{104}$Pd($p, \gamma$) + $^{105}$Pd($p, n$) reactions.}
\label{sigmaplus104105}
\renewcommand{\arraystretch}{1.1} % enlarge line spacing
\begin{ruledtabular}
  \begin{tabular}{ccc}
$E{\rm ^{eff}}$    & \multicolumn{2}{c}{Measured data ($^{104}$Pd($p, \gamma$) + $^{105}$Pd($p, n$))}  \\
                          & $\sigma^+$  & $S^+$ factor 	\\                       
 (MeV)					  				& (mbarn)			& (10$^{8}$ MeV b) 	\\
\hline
2.687$^{+0.012}_{-0.019}$ & 0.073$\pm$0.005 & 1.96$\pm$0.14 \\
2.933$^{+0.012}_{-0.018}$ & 0.185$\pm$0.013 & 1.67$\pm$0.12 \\
3.178$^{+0.012}_{-0.016}$ & 0.47$\pm$0.03 	& 1.61$\pm$0.11 \\
3.424$^{+0.013}_{-0.014}$ & 0.92$\pm$0.07 	& 1.35$\pm$0.10 \\
3.444$^{+0.013}_{-0.014}$ & 1.11$\pm$0.08 	& 1.52$\pm$0.12 \\
3.679$^{+0.025}_{-0.029}$ & 1.96$\pm$0.14 	& 1.30$\pm$0.09 \\
3.894$^{+0.026}_{-0.029}$ & 3.69$\pm$0.24 	& 1.34$\pm$0.09 \\
3.934$^{+0.027}_{-0.028}$ & 3.82$\pm$0.27 	& 1.24$\pm$0.09 \\
4.170$^{+0.026}_{-0.029}$ & 6.47$\pm$0.43 	& 1.16$\pm$0.08 \\
4.436$^{+0.027}_{-0.031}$ & 8.49$\pm$0.60 	& 0.95$\pm$0.07 \\
4.890$^{+0.028}_{-0.031}$ & 20.5$\pm$1.5 		& 0.79$\pm$0.06 \\
4.912$^{+0.029}_{-0.031}$ & 21.6$\pm$1.6 		& 0.80$\pm$0.06 \\
\end{tabular}
\end{ruledtabular}
\end{table}

%The measured composite cross sections have been decomposed with our experimental results for the $^{104}$Pd($p,\gamma$) cross section from enriched samples. The composite data and the resulting $^{105}$Pd($p, n$) cross sections are listed in .

\subsubsection{$^{104}$Pd($p, \gamma$)$^{105}$Ag measured with enriched samples}
The total $^{104}$Pd($p, \gamma$)$^{105}$Ag cross section was measured additionally with enriched $^{104}$Pd samples (Table~\ref{tab:104pg}). In Fig.~\ref{res-pd104105} our results are compared to the values of Ref.~\cite{spyr08} which are in very good agreement with the NON-SMOKER predictions. Our results are slightly lower, but follow the theoretical and experimental energy trend and -- with exception of the highest data point-- reproduce the theoretical values within the factor of two error band.

\begin{table}[!h]
\caption{\label{tab:104pg} Cross sections and $S$ factors for $^{104}$Pd($p, \gamma$)$^{105}$Ag from enriched samples.}
\renewcommand{\arraystretch}{1.2} % enlarge line spacing
\begin{ruledtabular}
\begin{tabular}{ccc}
$E{\rm ^{eff}}$    & Cross section   & $S$ factor \\
   (MeV)      			  & (mbarn) 		& (10$^7$ MeV b) \\
\hline
2.714$^{+0.010}_{-0.017}$ &  0.024$\pm$0.003  &  5.75$\pm$0.74 \\
2.963$^{+0.009}_{-0.014}$ &  0.091$\pm$0.011  &  7.24$\pm$0.89 \\ 
3.653$^{+0.023}_{-0.028}$ &  0.634$\pm$0.047  &  4.54$\pm$0.34 \\
3.991$^{+0.024}_{-0.028}$ &  1.654$\pm$0.124  &  4.64$\pm$0.32 \\
4.473$^{+0.026}_{-0.029}$ &  3.640$\pm$0.214  &  3.27$\pm$0.19 \\
5.035$^{+0.027}_{-0.031}$ &  6.104$\pm$0.394  &  1.80$\pm$0.12 \\
\end{tabular}
\end{ruledtabular}
\end{table}

\subsubsection{$^{105}$Pd($p, n$)$^{105}$Ag measured with enriched samples}
\begin{figure*}[!htb]
\includegraphics{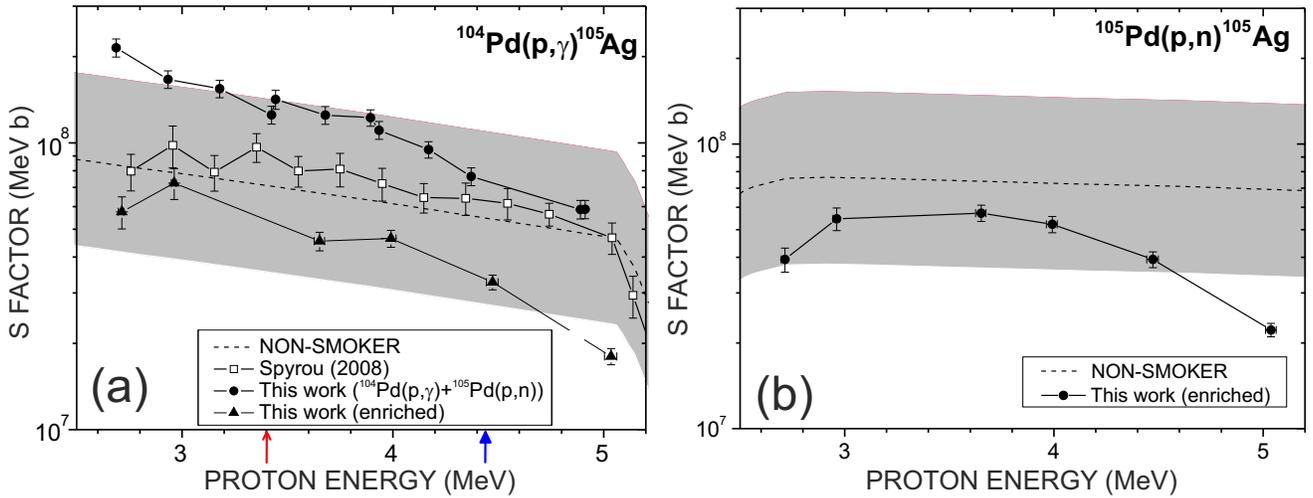}
\caption{\label{res-pd104105}(Color online) (a): $S$ factors for $^{104}$Pd($p, \gamma$)$^{105}$Ag. The data from Spyrou \cite{spyr08} is shown in comparison with our composite $S$ factor for $^{104}$Pd($p, \gamma$)+$^{105}$Pd($p, n$) and the results from the enriched samples. 
(b): $S$ factors for $^{105}$Pd($p, n$)$^{105}$Ag from the enriched samples. The predictions from NON-SMOKER \cite{rath00,rath01} (dashed line) are shown with a region of uncertainty of a factor of two. The thin and thick arrows indicate the upper ends of the respective Gamow windows for T=2 and 3~GK, respectively. For $^{105}$Pd($p, n$)$^{105}$Ag the limits are off scale, see Table~\ref{tab:gamow}.}
\end{figure*}

The $^{105}$Pd($p, n$)$^{105}$Ag reaction channel opens at 2.15~MeV. The total cross section could also be measured from enriched $^{105}$Pd samples via the above mentioned $\gamma$ transitions (Table~\ref{tab:105pn}). Fig.~\ref{res-pd104105} shows these results, which are again slightly lower than the NON-SMOKER prediction. With exception of the highest data point our results can reproduce the theoretical values within the factor of two region of uncertainty. Unfortunately there is no other data available for comparison.

\begin{table}[!h]
\caption{\label{tab:105pn} Cross sections and $S$ factors for 
$^{105}$Pd($p, n$)$^{105}$Ag from enriched samples.}
\renewcommand{\arraystretch}{1.2} % enlarge line spacing
\begin{ruledtabular}
\begin{tabular}{ccc}
$E{\rm ^{eff}}$    & Cross section   & $S$ factor \\
   (MeV)      			  & (mbarn) 		& (10$^7$ MeV b) \\
\hline
2.714$^{+0.011}_{-0.018}$ &  0.017$\pm$0.002  &  3.93$\pm$0.38 \\
2.961$^{+0.011}_{-0.017}$ &  0.068$\pm$0.006  &  5.47$\pm$0.50 \\
3.651$^{+0.024}_{-0.027}$ &  0.794$\pm$0.052  &  5.72$\pm$0.37 \\
3.991$^{+0.025}_{-0.027}$ &  1.861$\pm$0.122  &  5.22$\pm$0.34 \\
4.473$^{+0.027}_{-0.028}$ &  4.383$\pm$0.279  &  3.93$\pm$0.25 \\
5.038$^{+0.025}_{-0.027}$ &  7.571$\pm$0.410  &  2.22$\pm$0.12 \\
\end{tabular}
\end{ruledtabular}
\end{table}

\subsection{$^{105}$Pd($p, \gamma$)$^{106}$Ag$^{m}$ and $^{106}$Pd($p, n$)$^{106}$Ag$^{m}$}
The partial cross section to the ground state in $^{106}$Ag was
not accessible by the activations because of its short half-life of 
23.96~min. The EC decay of the 6$^+$ isomeric state with $t_{1/2}= 8.28$~d to 
$^{106}$Pd could be followed via the transitions at 451.0, 717.2, 
748.4, and 1045.8~keV. Measurements have been performed with natural 
and with enriched samples (Table~\ref{tab:abun}). 

\subsubsection{Cross sections measured with natural samples}
Above the threshold of the $^{106}$Pd($p, n$) reaction 
at 3.78 MeV the measured cross section represents the composite
of the partial cross sections for $^{105}$Pd($p, \gamma$)$^{106}$Ag$^{m}$
and $^{106}$Pd($p, n$)$^{106}$Ag$^{m}$, similar to the 
previous case. The results are listed in Table~\ref{tab:105106} and shown in Fig.~\ref{res-pd105106}. 

\begin{table}[!h]
\caption{Composite cross section and $S$ factor for the $^{105}$Pd($p, \gamma$)$^{106}$Ag$^{m}$ + $^{106}$Pd($p, n$)$^{106}$Ag$^{m}$ reactions. The horizontal line indicates where the $^{106}$Pd($p, n$) reaction starts to contribute. }
\label{tab:105106} 
\renewcommand{\arraystretch}{1.1} % enlarge line spacing
\begin{ruledtabular}
  \begin{tabular}{ccc}
$E{\rm ^{eff}}$   & \multicolumn{2}{c}{Measured data ($^{105}$Pd($p, \gamma$)+$^{106}$Pd($p, n$))} \\
                  & $\sigma^+$  					& $S^+$ factor 				\\                       
 (MeV)					  & ($\mu$barn)		 				& (10$^{6}$ MeV b)    \\
\hline
3.178$^{+0.013}_{-0.015}$ &	3.6$\pm$0.4		& 1.233$\pm$0.138 \\
3.424$^{+0.011}_{-0.016}$ & 5.0$\pm$0.6		& 0.735$\pm$0.093 \\
3.444$^{+0.024}_{-0.029}$ &	6.1$\pm$0.6		& 0.838$\pm$0.084 \\
3.679$^{+0.026}_{-0.029}$ &	9.6$\pm$1.6		& 0.640$\pm$0.107 \\
\hline
3.894$^{+0.028}_{-0.028}$ & 15.0$\pm$1.8 	& 0.546$\pm$0.064 \\
3.937$^{+0.025}_{-0.030}$ & 15.1$\pm$2.1 	& 0.488$\pm$0.069 \\
4.170$^{+0.027}_{-0.029}$ & 44.6$\pm$4.9 	& 0.802$\pm$0.088 \\
4.370$^{+0.028}_{-0.029}$ & 63.6$\pm$8.8 	& 0.718$\pm$0.100 \\
4.438$^{+0.025}_{-0.031}$ & 77.8$\pm$5.2 	& 0.753$\pm$0.050 \\
4.890$^{+0.030}_{-0.031}$ & 299$\pm$22 		& 1.152$\pm$0.083 \\
4.913$^{+0.028}_{-0.031}$ & 307$\pm$25		& 1.133$\pm$0.092 \\
\end{tabular}
\end{ruledtabular}
\end{table}

\subsubsection{$^{105}$Pd($p, \gamma$)$^{106}$Ag$^{m}$ measured with enriched samples}
\begin{figure*}[!htb]
\includegraphics{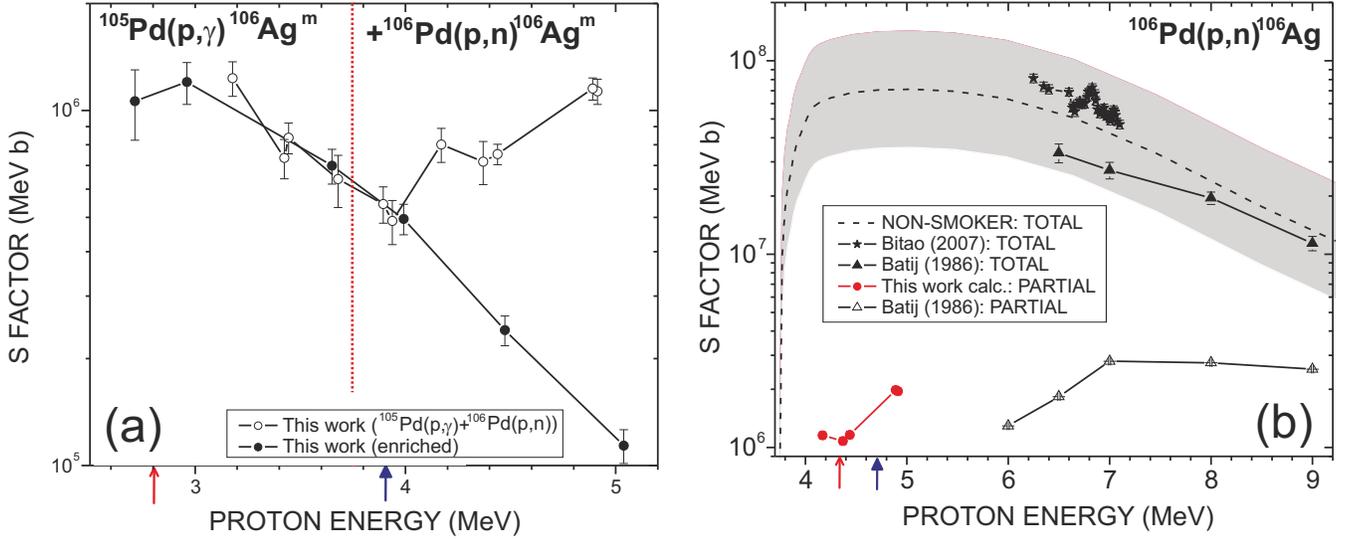}
\caption{\label{res-pd105106} (Color online) (a): $S$ factors for $^{105}$Pd($p,\gamma$)$^{106}$Ag$^{m}$. The $^{106}$Pd($p,n$)$^{106}$Ag$^{m}$ channel contributes above 3.78~MeV. (b): Decomposed $S$ factor for $^{106}$Pd($p,n$)$^{106}$Ag$^{m}$ in comparison with experimental data from \cite{BSR86} and \cite{bitao07}.
The $^{106}$Pd($p,n$)$^{106}$Ag threshold is indicated as horizontal red line. The prediction from NON-SMOKER \cite{rath00,rath01} (dashed line) is shown with a region of uncertainty of a factor of two. The thin and thick arrows indicate the upper ends of the respective Gamow windows for T=2 and 3~GK, respectively (see Table~\ref{tab:gamow}).}
\end{figure*}

The $^{105}$Pd($p, \gamma$)$^{106}$Ag$^{m}$ cross section was also measured with enriched $^{105}$Pd samples between $E_p$=2.7 and 5.0~MeV. The resulting $S$ factor (Table~\ref{tab:sigma105pg}) is shown in Fig.~\ref{res-pd105106} and agrees well with the data from the natural samples below the $^{106}$Pd($p, n$) threshold. 

The results obtained with the enriched samples could be used to decompose the cross section data in Table~\ref{tab:105106} to derive the cross section for the competing $^{106}$Pd($p, n$)$^{106}$Ag$^{m}$ channel (Table~\ref{tab:106pn}). Our deduced results are compared in Fig.~\ref{res-pd105106} with experimental data of Batij {\it et al.} \cite{BSR86}, which are reported at slightly higher energies. The results of Ref. \cite{BSR86} for the total cross section are higher than the values of Bitao {\it et al.} \cite{bitao07}.

\begin{table}[!h]
\caption{Cross sections and $S$ factors for the $^{105}$Pd($p, \gamma$)$^{106}$Ag$^{m}$ reaction measured with 
enriched samples.}\label{tab:sigma105pg}
\renewcommand{\arraystretch}{1.1} % enlarge line spacing
\begin{ruledtabular}
  \begin{tabular}{ccc}
$E{\rm ^{eff}}$    & Cross section  & $S$ factor          \\
  (MeV)  				  & ($\mu$barn)    & (10$^{6}$ MeV b) \\
\hline
2.714$^{+0.011}_{-0.018}$ & 0.44$\pm$0.10  & 1.063$\pm$0.238     \\
2.961$^{+0.011}_{-0.017}$ & 1.50$\pm$0.20  & 1.202$\pm$0.163 	 \\
3.462$^{+0.006}_{-0.009}$ & 3.99$\pm$0.65  & 0.518$\pm$0.084  	 \\
3.651$^{+0.024}_{-0.027}$ & 9.68$\pm$1.08  & 0.699$\pm$0.078 	 \\
3.991$^{+0.025}_{-0.027}$ & 17.6$\pm$1.7   & 0.495$\pm$0.049 	 \\
4.473$^{+0.027}_{-0.028}$ & 26.8$\pm$2.6   & 0.240$\pm$0.023 	 \\
5.038$^{+0.025}_{-0.027}$ & 38.7$\pm$4.2   & 0.114$\pm$0.012 	 \\
\end{tabular}
\end{ruledtabular}
\end{table}

\begin{table}
\caption{Decomposed cross sections and $S$ factors for the $^{106}$Pd($p,n$)$^{106}$Ag$^{m}$ reaction, obtained with our experimental results from $^{105}$Pd($p, \gamma$)$^{106}$Ag$^{m}$.}
\label{tab:106pn} 
\renewcommand{\arraystretch}{1.1} % enlarge line spacing
\begin{ruledtabular}
  \begin{tabular}{ccc}
$E{\rm ^{eff}}$    & Cross section & $S$ factor 				\\                       
 (MeV)					  & ($\mu$barn)		 	& (10$^{6}$ MeV b)    \\
\hline
4.170$^{+0.027}_{-0.029}$ & 63.9$\pm$7.0 	& 1.15$\pm$0.13 \\
4.370$^{+0.028}_{-0.029}$ & 95.4$\pm$13.3 & 1.08$\pm$0.15 \\
4.438$^{+0.025}_{-0.031}$ & 121$\pm$8 		& 1.17$\pm$0.08 \\
4.890$^{+0.030}_{-0.031}$ & 514$\pm$37 		& 1.98$\pm$0.14 \\
4.913$^{+0.028}_{-0.031}$ & 528$\pm$43		& 1.95$\pm$0.16 \\
\end{tabular}
\end{ruledtabular}
\end{table}

\subsection{$^{110}$Pd$(p, n)$$^{110}$Ag$^{m}$}

Activation of $^{110}$Pd provided only the partial ($p, n$) cross 
section to the isomeric state in $^{110}$Ag, which was measured via 
the $\gamma$-lines at 657.8, 763.9, 884.7 and 937.5~keV in the 
decay of $^{110}$Ag$^{m}$. The ground state of $^{110}$Ag
is too short-lived for the technique used in this work ($t_{1/2}=24$ s), and the 
NON-SMOKER calculation for the competing $^{110}$Pd($p, \gamma$) 
channel predicts a 1000 times lower cross section. Moreover, 
such events are difficult to detect because the strongest 
transition at 342.1~keV interferes with the strong 344.5~keV 
transition in $^{105}$Ag ($t_{1/2}$=41.29~d).    

\begin{figure}
\includegraphics{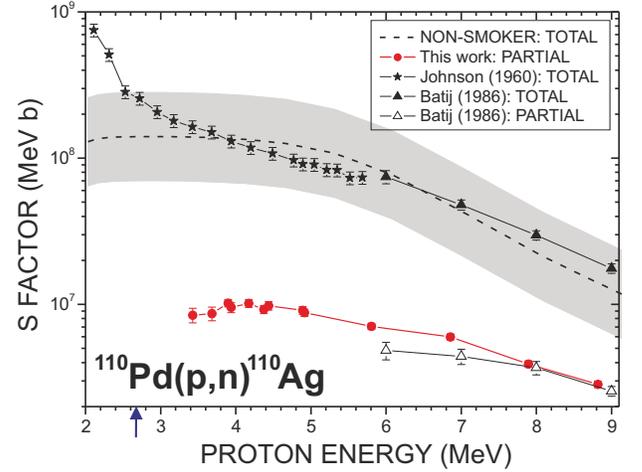}
\caption{\label{pd110-res}(Color online) $S$ factors for $^{110}$Pd$(p,n)$$^{110}$Ag compared to the results of Johnson {\it et al.} \cite{John60} (stars) and Batij {\it et al.} \cite{BSR86} (open and full triangles) for the total and partial cross sections. The NON-SMOKER predictions \cite{rath00,rath01} (dashed line) are plotted with a region of uncertainty of a factor of two. The blue arrow indicates the upper end of the respective Gamow window for T=3~GK (see Table~\ref{tab:gamow}).}
\end{figure}

The results for the cross section and the $S$ factor are 
summarized in Table~\ref{tab:110}. The comparison in 
Fig.~\ref{pd110-res} shows fair agreement (within $\sim25\%$) with 
the measurement of Batij {\it et al.} \cite{BSR86} between 6 
and 9 MeV. The total $(p, n)$ data of Batij {\it et al.} 
\cite{BSR86} and Johnson {\it et al.} \cite{John60} are included for 
illustrating the measured energy trends with respect to the 
NON-SMOKER prediction. While the present results for the partial 
data follow the predicted slope, the total $S$ factor of Johnson {\it et al.} exhibit increasing deviations  
toward lower energies. 

\begin{table}
\caption{\label{tab:110} Cross sections and $S$ factors for 
$^{110}$Pd($p, n$)$^{110}$Ag$^{m}$.}
\renewcommand{\arraystretch}{1.1} % enlarge line spacing
\begin{ruledtabular}
\begin{tabular}{ccc}
$E{\rm ^{eff}}$    & Cross section      & $S$ factor          \\
  (MeV)  				  & (mbarn) 		   & (10$^{6}$ MeV b) \\
\hline
3.426$^{+0.011}_{-0.016}$ &  0.058$\pm$0.007   &  8.45$\pm$0.98 	 \\
3.682$^{+0.025}_{-0.029}$ &  0.131$\pm$0.014   &  8.66$\pm$0.93 	 \\
3.898$^{+0.025}_{-0.029}$ &  0.282$\pm$0.016   &  10.2$\pm$0.58 	 \\
3.937$^{+0.026}_{-0.029}$ &  0.293$\pm$0.023   &  9.53$\pm$0.75 	 \\
4.172$^{+0.027}_{-0.029}$ &  0.566$\pm$0.035   &  10.2$\pm$0.63 	 \\
4.372$^{+0.028}_{-0.029}$ &  0.819$\pm$0.052   &  9.23$\pm$0.59 	 \\
4.439$^{+0.027}_{-0.031}$ &  1.01$\pm$0.07     &  9.77$\pm$0.67 	 \\
4.892$^{+0.030}_{-0.030}$ &  2.36$\pm$0.14     &  9.10$\pm$0.55		 \\
4.916$^{+0.027}_{-0.032}$ &  2.38$\pm$0.13     &  8.77$\pm$0.50 	 \\
5.805$^{+0.031}_{-0.034}$ &  8.29$\pm$0.38     &  7.08$\pm$0.33 	 \\
6.856$^{+0.035}_{-0.037}$ &  26.7$\pm$1.2      &  5.99$\pm$0.26 	 \\ 
7.896$^{+0.039}_{-0.040}$ &  49.1$\pm$2.1      &  3.90$\pm$0.17 	 \\ 
8.820$^{+0.041}_{-0.043}$ &  76.2$\pm$2.7      &  2.84$\pm$0.10 	 \\ 
\end{tabular}
\end{ruledtabular}
\end{table}

\section{Summary}

Extensive investigations of proton-induced reactions on Pd isotopes have been performed by means of the activation 
technique. The proton energy range between 2.7 and 8.8 MeV was chosen to cover relevant parts of the Gamow windows of the $p$ process. Total cross sections are reported for $^{102}$Pd($p, \gamma$)$^{103}$Ag, $^{104}$Pd($p, \gamma$)$^{105}$Ag, and $^{105}$Pd($p, n$)$^{105}$Ag. In addition, partial cross sections were determined for the reactions $^{104}$Pd($p, n$)$^{104}$Ag$^{g}$, $^{105}$Pd$(p,\gamma)$$^{106}$Ag$^{m}$, $^{106}$Pd($p, n$)$^{106}$Ag$^{m}$, and 
$^{110}$Pd($p, n$)$^{110}$Ag$^{m}$. Compared to previous experimental data the present results for $^{102}$Pd$(p, \gamma$)$^{103}$Ag are three times lower than reported by \"Ozkan {\it et al.} \cite{OMB02}, whereas fair agreement was found with the $^{104}$Pd($p,\gamma)$$^{105}$Ag data of Spyrou {\it et al.} \cite{spyr08}. For the partial $(p,n)$ cross sections measured in this work we find only in the case of $^{110}$Pd$(p,n)$$^{110}$Ag$^{m}$ a reasonable agreement with measured data of Batij {\it et al.} \cite{BSR86}. 

With respect to theory, the NON-SMOKER predictions \cite{rath00,rath01} for $^{102}$Pd($p, \gamma$), $^{104}$Pd$(p,\gamma)$, and $^{105}$Pd($p, n$) were confirmed. These results confirm also the overall good agreement of NON-SMOKER calculations for proton-induced reactions of isotopes between $^{70}$Ge and $^{209}$Bi in the energy range of the Gamow window of the $p$ process. From the 34 measured $(p,\gamma)$ reactions so far, only $^{98}$Ru, $^{114,116,119}$Sn, and $^{115}$In do not agree with NON-SMOKER within the factor of two uncertainty. In the cases of $^{98}$Ru and $^{115}$In this seems to be solely due to experimental problems. For $(p,n)$ reactions many more data are available within the Gamow window (datasets for 80 isotopes). A systematic comparison has not yet been carried out, but will be done in the $p$-process database of the "Karlsruhe Astrophysical Database of Nucleosynthesis in Stars" project \cite{SDP11}. 

The new reaction code SMARAGD \cite{SMA} will be used for future predictions of astrophysical reaction rates. For the reactions investigated here, the predictions of the current version SMARAGD v0.8.1s with standard settings are identical to the shown NON-SMOKER results.

\begin{acknowledgments}
We thank the operating team O. D\"ohr, H. Eggestein, T. Heldt, and M. Hoffmann for providing excellent beams at the PTB accelerators and W. Mannhardt for providing a second HPGe at the PTB. I.D. acknowledges the help of M. Guttmann (Institut f\"ur Materialforschung, IMF) und A. G\"ortzen (Institut f\"ur Nukleare Entsorgung, INE), both from Forschungszentrum Karlsruhe (now Karlsruhe Institute of Technology, KIT), with an electrolytic cell. I.D. is funded by the Helmholtz association via the Young Investigators project VH-NG-627. This work was partially supported by the Swiss National Science Foundation (grant 2000-113984/1). TR is supported by the European Commission within the FP7 ENSAR/THEXO project.
\end{acknowledgments}

%\bibliography{../dillmann-p}% Produces the bibliography via BibTeX.

\begin{thebibliography}{47}
\expandafter\ifx\csname natexlab\endcsname\relax\def\natexlab#1{#1}\fi
\expandafter\ifx\csname bibnamefont\endcsname\relax
  \def\bibnamefont#1{#1}\fi
\expandafter\ifx\csname bibfnamefont\endcsname\relax
  \def\bibfnamefont#1{#1}\fi
\expandafter\ifx\csname citenamefont\endcsname\relax
  \def\citenamefont#1{#1}\fi
\expandafter\ifx\csname url\endcsname\relax
  \def\url#1{\texttt{#1}}\fi
\expandafter\ifx\csname urlprefix\endcsname\relax\def\urlprefix{URL }\fi
\providecommand{\bibinfo}[2]{#2}
\providecommand{\eprint}[2][]{\url{#2}}

\bibitem[{\citenamefont{Burbidge et~al.}(1957)\citenamefont{Burbidge, Burbidge,
  Fowler, and Hoyle}}]{bbfh57}
\bibinfo{author}{\bibfnamefont{E.}~\bibnamefont{Burbidge}},
  \bibinfo{author}{\bibfnamefont{G.}~\bibnamefont{Burbidge}},
  \bibinfo{author}{\bibfnamefont{W.}~\bibnamefont{Fowler}}, \bibnamefont{and}
  \bibinfo{author}{\bibfnamefont{F.}~\bibnamefont{Hoyle}},
  \bibinfo{journal}{Rev. Mod. Phys.} \textbf{\bibinfo{volume}{29}},
  \bibinfo{pages}{547} (\bibinfo{year}{1957}).

\bibitem[{\citenamefont{Langanke and Wiescher}(2001)}]{lawi01}
\bibinfo{author}{\bibfnamefont{K.}~\bibnamefont{Langanke}} \bibnamefont{and}
  \bibinfo{author}{\bibfnamefont{M.}~\bibnamefont{Wiescher}},
  \bibinfo{journal}{Rep. Prog. Phys.} \textbf{\bibinfo{volume}{64}},
  \bibinfo{pages}{1657} (\bibinfo{year}{2001}).

\bibitem[{\citenamefont{Woosley and Howard}(1978)}]{woho78}
\bibinfo{author}{\bibfnamefont{S.}~\bibnamefont{Woosley}} \bibnamefont{and}
  \bibinfo{author}{\bibfnamefont{W.}~\bibnamefont{Howard}},
  \bibinfo{journal}{Astrophys. J. Suppl.} \textbf{\bibinfo{volume}{36}},
  \bibinfo{pages}{285} (\bibinfo{year}{1978}).

\bibitem[{\citenamefont{Woosley and Howard}(1990)}]{woho90}
\bibinfo{author}{\bibfnamefont{S.}~\bibnamefont{Woosley}} \bibnamefont{and}
  \bibinfo{author}{\bibfnamefont{W.}~\bibnamefont{Howard}},
  \bibinfo{journal}{Astrophys. J.} \textbf{\bibinfo{volume}{354}},
  \bibinfo{pages}{L21} (\bibinfo{year}{1990}).

\bibitem[{\citenamefont{Rayet et~al.}(1995)\citenamefont{Rayet, Arnould,
  Hashimoto, Prantzos, and Nomoto}}]{raar95}
\bibinfo{author}{\bibfnamefont{M.}~\bibnamefont{Rayet}},
  \bibinfo{author}{\bibfnamefont{M.}~\bibnamefont{Arnould}},
  \bibinfo{author}{\bibfnamefont{M.}~\bibnamefont{Hashimoto}},
  \bibinfo{author}{\bibfnamefont{N.}~\bibnamefont{Prantzos}}, \bibnamefont{and}
  \bibinfo{author}{\bibfnamefont{K.}~\bibnamefont{Nomoto}},
  \bibinfo{journal}{Astron. Astrophys.} \textbf{\bibinfo{volume}{298}},
  \bibinfo{pages}{517} (\bibinfo{year}{1995}).

\bibitem[{\citenamefont{Rauscher et~al.}(2002)\citenamefont{Rauscher, Heger,
  Hoffman, and Woosley}}]{rahe02}
\bibinfo{author}{\bibfnamefont{T.}~\bibnamefont{Rauscher}},
  \bibinfo{author}{\bibfnamefont{A.}~\bibnamefont{Heger}},
  \bibinfo{author}{\bibfnamefont{R.}~\bibnamefont{Hoffman}}, \bibnamefont{and}
  \bibinfo{author}{\bibfnamefont{S.}~\bibnamefont{Woosley}},
  \bibinfo{journal}{Astrophys. J.} \textbf{\bibinfo{volume}{576}},
  \bibinfo{pages}{323} (\bibinfo{year}{2002}).

\bibitem[{\citenamefont{Woosley et~al.}(1990)\citenamefont{Woosley, Hartmann,
  Hoffman, and Haxton}}]{WHH90}
\bibinfo{author}{\bibfnamefont{S.}~\bibnamefont{Woosley}},
  \bibinfo{author}{\bibfnamefont{D.}~\bibnamefont{Hartmann}},
  \bibinfo{author}{\bibfnamefont{R.}~\bibnamefont{Hoffman}}, \bibnamefont{and}
  \bibinfo{author}{\bibfnamefont{W.}~\bibnamefont{Haxton}},
  \bibinfo{journal}{Astrophys. J.} \textbf{\bibinfo{volume}{356}},
  \bibinfo{pages}{272} (\bibinfo{year}{1990}).

\bibitem[{\citenamefont{Heger et~al.}(2005)\citenamefont{Heger, Kolbe, Haxton,
  Langanke, Mart{\'i}nez-Pinedo, and Woosley}}]{HKH05}
\bibinfo{author}{\bibfnamefont{A.}~\bibnamefont{Heger}},
  \bibinfo{author}{\bibfnamefont{E.}~\bibnamefont{Kolbe}},
  \bibinfo{author}{\bibfnamefont{W.~C.} \bibnamefont{Haxton}},
  \bibinfo{author}{\bibfnamefont{K.}~\bibnamefont{Langanke}},
  \bibinfo{author}{\bibfnamefont{G.}~\bibnamefont{Mart{\'i}nez-Pinedo}},
  \bibnamefont{and} \bibinfo{author}{\bibfnamefont{S.~E.}
  \bibnamefont{Woosley}}, \bibinfo{journal}{Phys. Lett. B}
  \textbf{\bibinfo{volume}{606}}, \bibinfo{pages}{258} (\bibinfo{year}{2005}).

\bibitem[{\citenamefont{Rayet et~al.}(1990)\citenamefont{Rayet, Prantzos, and
  Arnould}}]{ray90}
\bibinfo{author}{\bibfnamefont{M.}~\bibnamefont{Rayet}},
  \bibinfo{author}{\bibfnamefont{N.}~\bibnamefont{Prantzos}}, \bibnamefont{and}
  \bibinfo{author}{\bibfnamefont{M.}~\bibnamefont{Arnould}},
  \bibinfo{journal}{Astron. Astrophys.} \textbf{\bibinfo{volume}{227}},
  \bibinfo{pages}{271} (\bibinfo{year}{1990}).

\bibitem[{\citenamefont{Rauscher et~al.}(1995)\citenamefont{Rauscher,
  Thielemann, and Oberhummer}}]{rau95}
\bibinfo{author}{\bibfnamefont{T.}~\bibnamefont{Rauscher}},
  \bibinfo{author}{\bibfnamefont{F.-K.} \bibnamefont{Thielemann}},
  \bibnamefont{and}
  \bibinfo{author}{\bibfnamefont{H.}~\bibnamefont{Oberhummer}},
  \bibinfo{journal}{Astrophys. J.} \textbf{\bibinfo{volume}{451}},
  \bibinfo{pages}{L37} (\bibinfo{year}{1995}).

\bibitem[{\citenamefont{Howard et~al.}(1991)\citenamefont{Howard, Meyer, and
  Woosley}}]{home91}
\bibinfo{author}{\bibfnamefont{W.}~\bibnamefont{Howard}},
  \bibinfo{author}{\bibfnamefont{B.}~\bibnamefont{Meyer}}, \bibnamefont{and}
  \bibinfo{author}{\bibfnamefont{S.}~\bibnamefont{Woosley}},
  \bibinfo{journal}{Astrophys. J.} \textbf{\bibinfo{volume}{373}},
  \bibinfo{pages}{L5} (\bibinfo{year}{1991}).

\bibitem[{\citenamefont{Goriely et~al.}(2002)\citenamefont{Goriely, Jos{\'e},
  Hernanz, Rayet, and Arnould}}]{GJH02}
\bibinfo{author}{\bibfnamefont{S.}~\bibnamefont{Goriely}},
  \bibinfo{author}{\bibfnamefont{J.}~\bibnamefont{Jos{\'e}}},
  \bibinfo{author}{\bibfnamefont{M.}~\bibnamefont{Hernanz}},
  \bibinfo{author}{\bibfnamefont{M.}~\bibnamefont{Rayet}}, \bibnamefont{and}
  \bibinfo{author}{\bibfnamefont{M.}~\bibnamefont{Arnould}},
  \bibinfo{journal}{Astron. Astrophys.} \textbf{\bibinfo{volume}{383}},
  \bibinfo{pages}{L27} (\bibinfo{year}{2002}).

\bibitem[{\citenamefont{Schatz et~al.}(1998)\citenamefont{Schatz, Aprahamian,
  G{\"o}rres, Wiescher, Rauscher, Rembges, Thielemann, Pfeiffer, M{\"o}ller,
  Herndl et~al.}}]{scha98}
\bibinfo{author}{\bibfnamefont{H.}~\bibnamefont{Schatz}},
  \bibinfo{author}{\bibfnamefont{A.}~\bibnamefont{Aprahamian}},
  \bibinfo{author}{\bibfnamefont{J.}~\bibnamefont{G{\"o}rres}},
  \bibinfo{author}{\bibfnamefont{M.}~\bibnamefont{Wiescher}},
  \bibinfo{author}{\bibfnamefont{T.}~\bibnamefont{Rauscher}},
  \bibinfo{author}{\bibfnamefont{J.}~\bibnamefont{Rembges}},
  \bibinfo{author}{\bibfnamefont{F.-K.} \bibnamefont{Thielemann}},
  \bibinfo{author}{\bibfnamefont{B.}~\bibnamefont{Pfeiffer}},
  \bibinfo{author}{\bibfnamefont{P.}~\bibnamefont{M{\"o}ller}},
  \bibinfo{author}{\bibfnamefont{H.}~\bibnamefont{Herndl}},
  \bibnamefont{et~al.}, \bibinfo{journal}{Phys. Rep.}
  \textbf{\bibinfo{volume}{294}}, \bibinfo{pages}{167} (\bibinfo{year}{1998}).

\bibitem[{\citenamefont{Schatz et~al.}(2001)\citenamefont{Schatz, Aprahamian,
  Barnard, Bildsten, Cumming, Ouellette, Rauscher, Thielemann, and
  Wiescher}}]{scha01}
\bibinfo{author}{\bibfnamefont{H.}~\bibnamefont{Schatz}},
  \bibinfo{author}{\bibfnamefont{A.}~\bibnamefont{Aprahamian}},
  \bibinfo{author}{\bibfnamefont{V.}~\bibnamefont{Barnard}},
  \bibinfo{author}{\bibfnamefont{L.}~\bibnamefont{Bildsten}},
  \bibinfo{author}{\bibfnamefont{A.}~\bibnamefont{Cumming}},
  \bibinfo{author}{\bibfnamefont{M.}~\bibnamefont{Ouellette}},
  \bibinfo{author}{\bibfnamefont{T.}~\bibnamefont{Rauscher}},
  \bibinfo{author}{\bibfnamefont{F.-K.} \bibnamefont{Thielemann}},
  \bibnamefont{and} \bibinfo{author}{\bibfnamefont{M.}~\bibnamefont{Wiescher}},
  \bibinfo{journal}{Phys. Rev. Lett.} \textbf{\bibinfo{volume}{86}},
  \bibinfo{pages}{3471} (\bibinfo{year}{2001}).

\bibitem[{\citenamefont{Fr{\"o}hlich et~al.}(2006)\citenamefont{Fr{\"o}hlich,
  Mart\'inez-Pinedo, Liebend{\"o}rfer, Thielemann, Bravo, Hix, Langanke, and
  Zinner}}]{FML06}
\bibinfo{author}{\bibfnamefont{C.}~\bibnamefont{Fr{\"o}hlich}},
  \bibinfo{author}{\bibfnamefont{G.}~\bibnamefont{Mart\'inez-Pinedo}},
  \bibinfo{author}{\bibfnamefont{M.}~\bibnamefont{Liebend{\"o}rfer}},
  \bibinfo{author}{\bibfnamefont{F.-K.} \bibnamefont{Thielemann}},
  \bibinfo{author}{\bibfnamefont{E.}~\bibnamefont{Bravo}},
  \bibinfo{author}{\bibfnamefont{W.~R.} \bibnamefont{Hix}},
  \bibinfo{author}{\bibfnamefont{K.}~\bibnamefont{Langanke}}, \bibnamefont{and}
  \bibinfo{author}{\bibfnamefont{N.~T.} \bibnamefont{Zinner}},
  \bibinfo{journal}{Phys. Rev. Lett.} \textbf{\bibinfo{volume}{96}},
  \bibinfo{pages}{142502} (\bibinfo{year}{2006}).

\bibitem[{\citenamefont{Dauphas et~al.}(2003)\citenamefont{Dauphas, Rauscher,
  Marty, and Reisberg}}]{Dau03}
\bibinfo{author}{\bibfnamefont{N.}~\bibnamefont{Dauphas}},
  \bibinfo{author}{\bibfnamefont{T.}~\bibnamefont{Rauscher}},
  \bibinfo{author}{\bibfnamefont{B.}~\bibnamefont{Marty}}, \bibnamefont{and}
  \bibinfo{author}{\bibfnamefont{L.}~\bibnamefont{Reisberg}},
  \bibinfo{journal}{Nucl. Phys.} \textbf{\bibinfo{volume}{A719}},
  \bibinfo{pages}{287c} (\bibinfo{year}{2003}).

\bibitem[{\citenamefont{Rauscher}(2010)}]{TR10}
\bibinfo{author}{\bibfnamefont{T.}~\bibnamefont{Rauscher}},
  \bibinfo{journal}{Phys. Rev. C} \textbf{\bibinfo{volume}{81}},
  \bibinfo{pages}{045807} (\bibinfo{year}{2010}).

\bibitem[{\citenamefont{Rauscher}(2006)}]{Rau06}
\bibinfo{author}{\bibfnamefont{T.}~\bibnamefont{Rauscher}},
  \bibinfo{journal}{Phys. Rev. C} \textbf{\bibinfo{volume}{73}},
  \bibinfo{pages}{015804} (\bibinfo{year}{2006}).

\bibitem[{\citenamefont{Kiss et~al.}(2008)\citenamefont{Kiss, Rauscher,
  Gy{\"u}rky, Simon, F{\"u}l{\"o}p, and Somorjai}}]{KRG08}
\bibinfo{author}{\bibfnamefont{G.~G.} \bibnamefont{Kiss}},
  \bibinfo{author}{\bibfnamefont{T.}~\bibnamefont{Rauscher}},
  \bibinfo{author}{\bibfnamefont{G.}~\bibnamefont{Gy{\"u}rky}},
  \bibinfo{author}{\bibfnamefont{A.}~\bibnamefont{Simon}},
  \bibinfo{author}{\bibfnamefont{Z.}~\bibnamefont{F{\"u}l{\"o}p}},
  \bibnamefont{and} \bibinfo{author}{\bibfnamefont{E.}~\bibnamefont{Somorjai}},
  \bibinfo{journal}{Phys. Rev. Lett.} \textbf{\bibinfo{volume}{101}},
  \bibinfo{pages}{191101} (\bibinfo{year}{2008}).

\bibitem[{\citenamefont{Rauscher et~al.}(2009)\citenamefont{Rauscher, Kiss,
  Gy{\"u}rky, Simon, F{\"u}l{\"o}p, and Somorjai}}]{RKG09}
\bibinfo{author}{\bibfnamefont{T.}~\bibnamefont{Rauscher}},
  \bibinfo{author}{\bibfnamefont{G.~G.} \bibnamefont{Kiss}},
  \bibinfo{author}{\bibfnamefont{G.}~\bibnamefont{Gy{\"u}rky}},
  \bibinfo{author}{\bibfnamefont{A.}~\bibnamefont{Simon}},
  \bibinfo{author}{\bibfnamefont{Z.}~\bibnamefont{F{\"u}l{\"o}p}},
  \bibnamefont{and} \bibinfo{author}{\bibfnamefont{E.}~\bibnamefont{Somorjai}},
  \bibinfo{journal}{Phys. Rev. C} \textbf{\bibinfo{volume}{80}},
  \bibinfo{pages}{035801} (\bibinfo{year}{2009}).

\bibitem[{\citenamefont{{Iliadis}}(2007)}]{Ili07}
\bibinfo{author}{\bibfnamefont{C.}~\bibnamefont{{Iliadis}}},
  \emph{\bibinfo{title}{{Nuclear Physics of Stars}}}
  (\bibinfo{publisher}{Wiley-VCH Verlag GmbH \& Co. KGaA, Weinheim/ Germany},
  \bibinfo{year}{2007}).

\bibitem[{\citenamefont{Utsunomiya et~al.}(2006)\citenamefont{Utsunomiya, Mohr,
  Zilges, and Rayet}}]{UMZ06}
\bibinfo{author}{\bibfnamefont{H.}~\bibnamefont{Utsunomiya}},
  \bibinfo{author}{\bibfnamefont{P.}~\bibnamefont{Mohr}},
  \bibinfo{author}{\bibfnamefont{A.}~\bibnamefont{Zilges}}, \bibnamefont{and}
  \bibinfo{author}{\bibfnamefont{M.}~\bibnamefont{Rayet}},
  \bibinfo{journal}{Nucl. Phys.} \textbf{\bibinfo{volume}{A777}},
  \bibinfo{pages}{459} (\bibinfo{year}{2006}).

\bibitem[{\citenamefont{Rapp et~al.}(2006)\citenamefont{Rapp, G{\"o}rres,
  Wiescher, Schatz, and K{\"a}ppeler}}]{RGW06}
\bibinfo{author}{\bibfnamefont{W.}~\bibnamefont{Rapp}},
  \bibinfo{author}{\bibfnamefont{J.}~\bibnamefont{G{\"o}rres}},
  \bibinfo{author}{\bibfnamefont{M.}~\bibnamefont{Wiescher}},
  \bibinfo{author}{\bibfnamefont{H.}~\bibnamefont{Schatz}}, \bibnamefont{and}
  \bibinfo{author}{\bibfnamefont{F.}~\bibnamefont{K{\"a}ppeler}},
  \bibinfo{journal}{Ap. J.} \textbf{\bibinfo{volume}{653}},
  \bibinfo{pages}{474} (\bibinfo{year}{2006}).

\bibitem[{\citenamefont{Brede et~al.}(1980)\citenamefont{Brede, Cosack, Dietze,
  Gumpert, Guldbakke, Jahr, Kutscha, Schlegel-Bickmann, and
  Sch{\"o}lermann}}]{BCD80}
\bibinfo{author}{\bibfnamefont{H.~J.} \bibnamefont{Brede}},
  \bibinfo{author}{\bibfnamefont{M.}~\bibnamefont{Cosack}},
  \bibinfo{author}{\bibfnamefont{G.}~\bibnamefont{Dietze}},
  \bibinfo{author}{\bibfnamefont{H.}~\bibnamefont{Gumpert}},
  \bibinfo{author}{\bibfnamefont{S.}~\bibnamefont{Guldbakke}},
  \bibinfo{author}{\bibfnamefont{R.}~\bibnamefont{Jahr}},
  \bibinfo{author}{\bibfnamefont{M.}~\bibnamefont{Kutscha}},
  \bibinfo{author}{\bibfnamefont{D.}~\bibnamefont{Schlegel-Bickmann}},
  \bibnamefont{and}
  \bibinfo{author}{\bibfnamefont{H.}~\bibnamefont{Sch{\"o}lermann}},
  \bibinfo{journal}{Nucl. Instr. Meth.} \textbf{\bibinfo{volume}{169}},
  \bibinfo{pages}{349} (\bibinfo{year}{1980}).

\bibitem[{\citenamefont{Trautmann and Folger}(1989)}]{TF89}
\bibinfo{author}{\bibfnamefont{N.}~\bibnamefont{Trautmann}} \bibnamefont{and}
  \bibinfo{author}{\bibfnamefont{H.}~\bibnamefont{Folger}},
  \bibinfo{journal}{Nucl. Instr. Meth. A} \textbf{\bibinfo{volume}{282}},
  \bibinfo{pages}{102} (\bibinfo{year}{1989}).

\bibitem[{\citenamefont{De~Laeter et~al.}(2003)\citenamefont{De~Laeter,
  B{\"o}hlke, de~Bievre, Hidaka, Peiser, Rosman, and Taylor}}]{iupac03}
\bibinfo{author}{\bibfnamefont{J.}~\bibnamefont{De~Laeter}},
  \bibinfo{author}{\bibfnamefont{J.}~\bibnamefont{B{\"o}hlke}},
  \bibinfo{author}{\bibfnamefont{P.}~\bibnamefont{de~Bievre}},
  \bibinfo{author}{\bibfnamefont{H.}~\bibnamefont{Hidaka}},
  \bibinfo{author}{\bibfnamefont{H.}~\bibnamefont{Peiser}},
  \bibinfo{author}{\bibfnamefont{K.}~\bibnamefont{Rosman}}, \bibnamefont{and}
  \bibinfo{author}{\bibfnamefont{P.}~\bibnamefont{Taylor}},
  \bibinfo{journal}{Pure and Appl. Chem.} \textbf{\bibinfo{volume}{75}},
  \bibinfo{pages}{683} (\bibinfo{year}{2003}).

\bibitem[{\citenamefont{B{\"o}ttger}(2002)}]{Boe02}
\bibinfo{author}{\bibfnamefont{R.}~\bibnamefont{B{\"o}ttger}},
  \bibinfo{journal}{private communication}  (\bibinfo{year}{2002}).

\bibitem[{\citenamefont{Ziegler and Biersack}(2003)}]{SRIM03}
\bibinfo{author}{\bibfnamefont{J.}~\bibnamefont{Ziegler}} \bibnamefont{and}
  \bibinfo{author}{\bibfnamefont{J.}~\bibnamefont{Biersack}},
  \bibinfo{journal}{''The stopping and range of ions in matter'', SRIM-2003.26,
  http://www.srim.com}  (\bibinfo{year}{2003}).

\bibitem[{\citenamefont{Beer and K{\"a}ppeler}(1980)}]{beer80}
\bibinfo{author}{\bibfnamefont{H.}~\bibnamefont{Beer}} \bibnamefont{and}
  \bibinfo{author}{\bibfnamefont{F.}~\bibnamefont{K{\"a}ppeler}},
  \bibinfo{journal}{Phys. Rev. C} \textbf{\bibinfo{volume}{21}},
  \bibinfo{pages}{534} (\bibinfo{year}{1980}).

\bibitem[{\citenamefont{de~Frenne and Jacobs}(2001)}]{nds103}
\bibinfo{author}{\bibfnamefont{D.}~\bibnamefont{de~Frenne}} \bibnamefont{and}
  \bibinfo{author}{\bibfnamefont{E.}~\bibnamefont{Jacobs}},
  \bibinfo{journal}{Nucl. Data Sheets} \textbf{\bibinfo{volume}{93}},
  \bibinfo{pages}{447} (\bibinfo{year}{2001}).

\bibitem[{\citenamefont{Blachot}(2007)}]{nds104}
\bibinfo{author}{\bibfnamefont{J.}~\bibnamefont{Blachot}},
  \bibinfo{journal}{Nucl. Data Sheets} \textbf{\bibinfo{volume}{108}},
  \bibinfo{pages}{2035} (\bibinfo{year}{2007}).

\bibitem[{\citenamefont{de~Frenne and Jacobs}(2005)}]{nds105}
\bibinfo{author}{\bibfnamefont{D.}~\bibnamefont{de~Frenne}} \bibnamefont{and}
  \bibinfo{author}{\bibfnamefont{E.}~\bibnamefont{Jacobs}},
  \bibinfo{journal}{Nucl. Data Sheets} \textbf{\bibinfo{volume}{105}},
  \bibinfo{pages}{775} (\bibinfo{year}{2005}).

\bibitem[{\citenamefont{de~Frenne and Negret}(2008)}]{nds106a}
\bibinfo{author}{\bibfnamefont{D.}~\bibnamefont{de~Frenne}} \bibnamefont{and}
  \bibinfo{author}{\bibfnamefont{A.}~\bibnamefont{Negret}},
  \bibinfo{journal}{Nucl. Data Sheets} \textbf{\bibinfo{volume}{109}},
  \bibinfo{pages}{943} (\bibinfo{year}{2008}).

\bibitem[{\citenamefont{de~Frenne and Jacobs}(2000)}]{nds110}
\bibinfo{author}{\bibfnamefont{D.}~\bibnamefont{de~Frenne}} \bibnamefont{and}
  \bibinfo{author}{\bibfnamefont{E.}~\bibnamefont{Jacobs}},
  \bibinfo{journal}{Nucl. Data Sheets} \textbf{\bibinfo{volume}{89}},
  \bibinfo{pages}{481} (\bibinfo{year}{2000}).

\bibitem[{\citenamefont{Debertin and Helmer}(1989)}]{DeHe}
\bibinfo{author}{\bibfnamefont{K.}~\bibnamefont{Debertin}} \bibnamefont{and}
  \bibinfo{author}{\bibfnamefont{R.}~\bibnamefont{Helmer}},
  \emph{\bibinfo{title}{{Gamma- and X-Ray Spectrometry With Semiconductor
  Detectors}}} (\bibinfo{publisher}{North Holland, ISBN 0444871071},
  \bibinfo{year}{1989}).

\bibitem[{\citenamefont{Rauscher and Thielemann}(2000)}]{rath00}
\bibinfo{author}{\bibfnamefont{T.}~\bibnamefont{Rauscher}} \bibnamefont{and}
  \bibinfo{author}{\bibfnamefont{F.-K.} \bibnamefont{Thielemann}},
  \bibinfo{journal}{At. Data Nucl. Data Tables} \textbf{\bibinfo{volume}{75}},
  \bibinfo{pages}{1} (\bibinfo{year}{2000}).

\bibitem[{\citenamefont{Rauscher and Thielemann}(2001)}]{rath01}
\bibinfo{author}{\bibfnamefont{T.}~\bibnamefont{Rauscher}} \bibnamefont{and}
  \bibinfo{author}{\bibfnamefont{F.-K.} \bibnamefont{Thielemann}},
  \bibinfo{journal}{At. Data Nucl. Data Tables} \textbf{\bibinfo{volume}{79}},
  \bibinfo{pages}{47} (\bibinfo{year}{2001}).

\bibitem[{\citenamefont{{\"O}zkan et~al.}(2002)\citenamefont{{\"O}zkan, Murphy,
  Boyd, Cole, Famiano, G{\"u}ray, Howard, Sahin, Zach, deHaan et~al.}}]{OMB02}
\bibinfo{author}{\bibfnamefont{N.}~\bibnamefont{{\"O}zkan}},
  \bibinfo{author}{\bibfnamefont{A.}~\bibnamefont{Murphy}},
  \bibinfo{author}{\bibfnamefont{R.}~\bibnamefont{Boyd}},
  \bibinfo{author}{\bibfnamefont{A.}~\bibnamefont{Cole}},
  \bibinfo{author}{\bibfnamefont{M.}~\bibnamefont{Famiano}},
  \bibinfo{author}{\bibfnamefont{R.}~\bibnamefont{G{\"u}ray}},
  \bibinfo{author}{\bibfnamefont{M.}~\bibnamefont{Howard}},
  \bibinfo{author}{\bibfnamefont{L.}~\bibnamefont{Sahin}},
  \bibinfo{author}{\bibfnamefont{J.}~\bibnamefont{Zach}},
  \bibinfo{author}{\bibfnamefont{R.}~\bibnamefont{deHaan}},
  \bibnamefont{et~al.}, \bibinfo{journal}{Nucl. Phys. A}
  \textbf{\bibinfo{volume}{710}}, \bibinfo{pages}{469} (\bibinfo{year}{2002}).

\bibitem[{\citenamefont{Batij et~al.}(1986)\citenamefont{Batij, Skakun,
  Rakivnenko, and Rastrepin}}]{BSR86}
\bibinfo{author}{\bibfnamefont{V.}~\bibnamefont{Batij}},
  \bibinfo{author}{\bibfnamefont{E.}~\bibnamefont{Skakun}},
  \bibinfo{author}{\bibfnamefont{J.}~\bibnamefont{Rakivnenko}},
  \bibnamefont{and}
  \bibinfo{author}{\bibfnamefont{O.}~\bibnamefont{Rastrepin}},
  \bibinfo{journal}{Proc. 36th Conference on \textit{Nucl. Spectr. and Nucl.
  Struc.}, Kharkov, USSR} p. \bibinfo{pages}{280} (\bibinfo{year}{1986}).

\bibitem[{\citenamefont{Lejeune}(1980)}]{lej80}
\bibinfo{author}{\bibfnamefont{A.}~\bibnamefont{Lejeune}},
  \bibinfo{journal}{Phys. Rev. C} \textbf{\bibinfo{volume}{21}},
  \bibinfo{pages}{1107} (\bibinfo{year}{1980}).

\bibitem[{\citenamefont{{E}xperimental {N}uclear {R}eaction~{D}ata
  EXFOR}(2009)}]{exfor}
\bibinfo{author}{\bibnamefont{{E}xperimental {N}uclear {R}eaction~{D}ata
  EXFOR}}, \bibinfo{journal}{Online: http://www-nds.iaea.org/exfor/exfor.htm}
  (\bibinfo{year}{2009}).

\bibitem[{\citenamefont{Bitao et~al.}(1998)\citenamefont{Bitao, Zarubin, and
  Juralev}}]{BZJ98}
\bibinfo{author}{\bibfnamefont{H.}~\bibnamefont{Bitao}},
  \bibinfo{author}{\bibfnamefont{P.}~\bibnamefont{Zarubin}}, \bibnamefont{and}
  \bibinfo{author}{\bibfnamefont{U.}~\bibnamefont{Juralev}},
  \bibinfo{journal}{Eur. Phys. J. A} \textbf{\bibinfo{volume}{2}},
  \bibinfo{pages}{143} (\bibinfo{year}{1998}).

\bibitem[{\citenamefont{Spyrou et~al.}(2008)\citenamefont{Spyrou, Lagoyannis,
  Demetriou, Harissopulos, and Becker}}]{spyr08}
\bibinfo{author}{\bibfnamefont{A.}~\bibnamefont{Spyrou}},
  \bibinfo{author}{\bibfnamefont{A.}~\bibnamefont{Lagoyannis}},
  \bibinfo{author}{\bibfnamefont{P.}~\bibnamefont{Demetriou}},
  \bibinfo{author}{\bibfnamefont{S.}~\bibnamefont{Harissopulos}},
  \bibnamefont{and} \bibinfo{author}{\bibfnamefont{H.}~\bibnamefont{Becker}},
  \bibinfo{journal}{Phys. Rev. C} \textbf{\bibinfo{volume}{77}},
  \bibinfo{pages}{065801} (\bibinfo{year}{2008}).

\bibitem[{\citenamefont{Bi-Tao et~al.}(2007)\citenamefont{Bi-Tao, Zarubin, and
  Juralev}}]{bitao07}
\bibinfo{author}{\bibfnamefont{H.}~\bibnamefont{Bi-Tao}},
  \bibinfo{author}{\bibfnamefont{P.}~\bibnamefont{Zarubin}}, \bibnamefont{and}
  \bibinfo{author}{\bibfnamefont{U.}~\bibnamefont{Juralev}},
  \bibinfo{journal}{Chinese Physics} \textbf{\bibinfo{volume}{16}},
  \bibinfo{pages}{989} (\bibinfo{year}{2007}).

\bibitem[{\citenamefont{Johnson et~al.}(1960)\citenamefont{Johnson, Galonsky,
  and Inskeep}}]{John60}
\bibinfo{author}{\bibfnamefont{C.}~\bibnamefont{Johnson}},
  \bibinfo{author}{\bibfnamefont{A.}~\bibnamefont{Galonsky}}, \bibnamefont{and}
  \bibinfo{author}{\bibfnamefont{C.}~\bibnamefont{Inskeep}},
  \bibinfo{journal}{Oak Ridge National Laboratory Reports ORNL-2910}
  p.~\bibinfo{pages}{25} (\bibinfo{year}{1960}).

\bibitem[{\citenamefont{Sz{\"u}cs et~al.}(2011)\citenamefont{Sz{\"u}cs,
  Dillmann, Plag, and F{\"u}l{\"o}p}}]{SDP11}
\bibinfo{author}{\bibfnamefont{T.}~\bibnamefont{Sz{\"u}cs}},
  \bibinfo{author}{\bibfnamefont{I.}~\bibnamefont{Dillmann}},
  \bibinfo{author}{\bibfnamefont{R.}~\bibnamefont{Plag}}, \bibnamefont{and}
  \bibinfo{author}{\bibfnamefont{Z.}~\bibnamefont{F{\"u}l{\"o}p}},
  \bibinfo{journal}{11th Symposium on Nuclei in the Cosmos, Proceedings of
  Science, PoS(NIC XI)247}  (\bibinfo{year}{2011}); Online: www.kadonis.org.

\bibitem[{\citenamefont{Rauscher}(2011)}]{SMA}
\bibinfo{author}{\bibfnamefont{T.}~\bibnamefont{Rauscher}},
  \bibinfo{journal}{Int. J. Mod. Phys. E} \textbf{\bibinfo{volume}{20}},
  \bibinfo{pages}{1071} (\bibinfo{year}{2011}).

\end{thebibliography}

\end{document}